\documentclass[conference,compsoc]{IEEEtran}
\usepackage[usenames,dvipsnames,svgnames,x11names]{xcolor}
\newcommand{\ysnote}[1]{ {\textcolor{magenta} { ***Yogesh: #1 }}} 
\newcommand{\drnote}[1]{ {\textcolor{orange} { ***Dreamer: #1 }}}
\newcommand{\ysnoted}[1]{} 

\usepackage[normalem]{ulem} 

\usepackage[nocompress]{cite}
\usepackage{lipsum}
\let\OLDthebibliography\thebibliography
\renewcommand\thebibliography[1]{
  \OLDthebibliography{#1}
  \setlength{\parskip}{0pt}
  \setlength{\itemsep}{0pt plus 0.3ex}
}

\usepackage{listings}

\lstdefinelanguage{XML}
{
basicstyle=\ttfamily\footnotesize,
  morestring=[b]",
  moredelim=[s][\bfseries\color{Maroon}]{<}{\ },
  moredelim=[s][\bfseries\color{Maroon}]{</}{>},
  moredelim=[l][\bfseries\color{Maroon}]{/>},
  moredelim=[l][\bfseries\color{Maroon}]{>},
  morecomment=[s]{<?}{?>},
  morecomment=[s]{<!--}{-->},
  commentstyle=\color{gray},
  stringstyle=\color{blue},
  identifierstyle=\color{red}
}
\usepackage{moreverb}

\usepackage[nounderscore]{syntax}

\usepackage[pdftex]{graphicx}
\graphicspath{{./figures/}}
\DeclareGraphicsExtensions{.pdf}
\usepackage[cmex10]{amsmath}
\usepackage{amssymb}
\usepackage{mathtools}
\usepackage{amsthm}
\usepackage{amsfonts}
\newtheorem{cons}{Constraint}
\usepackage{subfig} 
\usepackage{algorithmicx}
\usepackage{algpseudocode}
\usepackage[ruled]{algorithm}
\definecolor{light-gray}{gray}{0.75}
\algrenewcommand{\algorithmiccomment}[1]{\hskip3em{{\footnotesize \textcolor{light-gray}{$\blacktriangleright$}}} #1}
\usepackage{multirow} 
\usepackage{rotating} 
\usepackage{booktabs} 
\usepackage{colortbl} 
\usepackage{tablefootnote} 

\usepackage[pdftex,colorlinks=true,urlcolor=blue,citecolor=blue]{hyperref}

\usepackage{xspace}

\usepackage{enumitem}
\newcommand{\para}[1]{\noindent\textbf{#1.~}}

\usepackage{blindtext}

\begin{document}
%
\title{Adaptive Energy-aware Scheduling of Dynamic Event Analytics\\ across Edge and Cloud Resources}

\IEEEoverridecommandlockouts
\author{\IEEEauthorblockN{Rajrup Ghosh*\thanks{*This work was done as a part of research at the Indian Institute of Science, Bangalore. The author is currently affiliated with Samsung R\&D Institute, Bangalore.}, Siva Prakash Reddy Komma and Yogesh Simmhan}
\IEEEauthorblockA{Computational and Data Sciences, Indian Institute of Science, Bangalore, India\\
Email: rajrup.withbestwishes@gmail.com, sivaprakash@iisc.ac.in, simmhan@iisc.ac.in}
}


%


\maketitle

\begin{abstract}
The growing deployment of sensors as part of \emph{Internet of Things (IoT)} is generating thousands of event streams. \emph{Complex Event Processing (CEP)} queries offer a useful paradigm for rapid decision-making over such data sources. While often centralized in the Cloud, the deployment of capable edge devices on the field motivates the need for cooperative event analytics that span \emph{Edge and Cloud computing}. Here, we identify a novel problem of query placement on edge and Cloud resources for dynamically arriving and departing analytic dataflows. We define this as an optimization problem to minimize the total makespan for all event analytics, while meeting energy and compute constraints of the resources. We propose $4$ adaptive heuristics and $3$ rebalancing strategies for such dynamic dataflows, and validate them using detailed simulations for $100-1000$ edge devices and VMs. The results show that our heuristics offer $\mathcal{O}(seconds)$ planning time, give a valid and high quality solution in all cases, and reduce the number of query migrations. Furthermore, rebalance  strategies when applied in these heuristics have significantly reduced the makespan by around $20-25\%$.
\end{abstract}


%
\IEEEpeerreviewmaketitle

\section{Introduction}
Internet of Things (IoT) is a distributed systems paradigm where sensors and actuators are connected through communication channels with the wider Internet to help observe and control large and complex physical systems. IoT manifests itself in various application domains, such as \emph{Industrial IoT}~\cite{bloem2014fourth} 
and \emph{Smart utilities}. E.g., a city utility may monitor the power grid and water network to identify power outages and water leaks or to predict near-term demand, based on real-time consumer and network observations~\cite{simmhan:cise:2012}.

Event analytics form a key aspect of this ``smartness'', and translates the observations from numerous sensors into actionable intelligence~\cite{strohbach2015towards}. \emph{Complex Event Processing (CEP)} is one form of event analytics where continuous queries are defined over one or more event streams to detect patterns of interest~\cite{cep-survey}. Such patterns can include \emph{filters} on properties in the events, \emph{aggregation} over a time or count window or events, or a \emph{sequence} of events that match a trend. A CEP query is similar to an SQL query defined over unbounded and transient stream(s) 
of tuples, and that generates one or more output event streams. CEP query engines such as \emph{Apache Edgent}~\cite{edgent} and \emph{WSO2 Siddhi}~\cite{siddhi11}, perform these queries with low latency and have been applied to IoT domains for online decision-making~\cite{debs-challenge-plug,strohbach2015towards}. The queries themselves can be chained and composed into arbitrary dataflows, each of which we refer to as an \emph{event analytic}, to capture complex situations and their event-driven responses.

Event analytics are typically performed centrally on Cloud resources. However, IoT deployments offer captive access to gateway devices with non-trivial computing capacity~\cite{varshney:icfec,kumar:icml:2017}. E.g., a \emph{Raspberry Pi 2} device, often used as an IoT gateway, has 4-core ARM 64-bit CPU and sells for $US\$~35$, but has about $\frac{1}{3}^{rd}$ the compute capacity for CEP analytics as a 4-core Intel Xeon CPU on a \emph{Microsoft Azure VM} that costs $\approx US\$~150/month$~\cite{ghosh:tcps:2017}. An IoT infrastructure may have hundreds of such ``free'' edge devices available. 

Besides the cost benefits, edge devices also have lower network latency from the sensor event source to the device, in comparison to the Cloud data center~\cite{satya:comm:2015}. At the same time, Cloud resources are more reliable, and have seemingly infinite on-demand capacity for a fee. Edges suffer from limited resource capacity on individual devices, and often have energy constraints, e.g., based on the recharge cycle for a solar-powered Pi. As a result, using \emph{edge-computing} as a first class resource to complement Cloud computing is essential for IoT applications~\cite{epema:edge}.

\textbf{Problem.} Given an IoT deployment of sensors and edge devices, we have a set of users and applications who wish to perform event analytics over the sensor event streams using the edge and Cloud resources that are available. These event analytics themselves are transient, being registered and active for a certain period of time before  user deregister them once their interest passes. As a result, we have a dynamic situation where the analytics arrive and depart, will be sharing the same set of edge and Cloud resources, and need to perform their CEP queries with low latency. 
\emph{We propose adaptive scheduling strategies to place the CEP queries from the dynamic analytic dataflows onto the distributed edge and Cloud resources to minimize the total makespan for all dataflows, while addressing the compute and network constraints of the resources, the energy constraints of the edge devices, and interference between queries on the same resource.} This problem is motivated by real-world concerns IoT deployments for smart water and power management~\cite{smartx,simmhan:cise:2012}, with city-scale applications as well~\cite{amrutur:lightpole}.

In our prior work~\cite{ghosh:tcps:2017}, we have considered \emph{static} scheduling of a \emph{single} CEP dataflow on a set of edge and Cloud resources while meeting the energy and compute constraints, which was solved using a \emph{Genetic Algorithm (GA) meta-heuristic}. Here, we consider multiple analytic dataflows that arrive and depart the system, propose novel heuristics and rebalancing strategies for the deployed dataflows, and evaluate these strategies against the earlier GA approach.

We make the following specific contributions here:
\begin{enumerate}
\item We motivate the need to schedule event analytics on edge and Cloud resources (\S~\ref{sec:motive}), and \emph{formalize the optimization problem} and constraints for adaptive placement of dataflows that dynamically arrive/depart (\S~\ref{sec:problem}).
\item We propose \emph{novel heuristics and rebalancing strategies} to solve the above problem, besides extending a prior GA-based approach for a dynamic scenario (\S~\ref{sec:solution}).
\item We \emph{validate the scheduling strategies} for their quality (latency and stabilization time) and performance (time complexity) using $39$ real CEP dataflows on $100-1000$ edge devices and VMs, with arrival and departure modeled as random walk and Poisson distributions (\S~\ref{sec:results}).
\end{enumerate}
In addition, we also review related literature (\S~\ref{sec:related}) and summarize our conclusions (\S~\ref{sec:conclusions}).

\section{Motivation}
\label{sec:motive}
Campus and community scale IoT testbeds, as well as a few city-scale deployments, are coming online~\cite{simmhan:cise:2012,yannuzzi-2017,amrutur:lightpole}. One such example is the \emph{Smart Campus Project}~\cite{smartx} which is deploying an IoT fabric of sensors, actuators and gateway devices across the IISc university campus in Bangalore. This will help understand practical and research challenges on Smart City platforms~\cite{mishra:iotn:2015}. The campus is spread across $1.6$km$^2$ with over 10,000 students, staff and faculty, and 50 buildings, that is representative of a small township. The deployment initially targets \emph{smart water management}, with water level and quality sensors and flowmeters deployed, along with actuators to control pumping and valve operations. Event streams from hundreds of sensor observations flow through Raspberry Pi gateway devices that sit between the sensors, connected through ZigBee/LoRA wireless protocols, and the campus backbone. Event analytics drive decision-making such as turning on and off pumps when the water level is low or high, and sending alerts when water quality drops or leakages in the network is detected. 

Numerous CEP queries help with online processing of these real-time streams to detect such situations of interest. E.g., the following queries over water quality and level sensors, sampled every $5~mins$, detect when the chlorine level drops below a safety threshold~\cite{who-chlorine}, the water level drops rapidly over $15~mins$ indicating leakage, and report the average sliding usage over $60~mins$. They are described using the Siddhi CEP engine's query model, and represent \emph{filter, sequence,} and \emph{aggregate} query types.

{\footnotesize
\begin{lstlisting}
$\circ$  |\textbf{FROM}| qltyStm[clMgL < 0.2]
  |\textbf{SELECT}| clMgL |\textbf{INSERT INTO}| qltyAlertStm; |\vspace{0.1cm}|
$\circ$  |\textbf{FROM EVERY}| l1=lvlStm,l2=lvlStm[l1.htCm-l2.hCm>25],
   l3=lvlStm[l2.htCm-l3.htCm>25]
  |\textbf{SELECT}| l1.htCm,l3.htCm |\textbf{INSERT INTO}| leakAlertStm; |\vspace{0.1cm}|
$\circ$  |\textbf{FROM}| lvlStm |\emph{\#window.length(12)}|
  |\textbf{SELECT}| |\emph{avg}|(htCm) |\textbf{as}| avgHt |\textbf{INSERT INTO}| avgLvlStm;
\end{lstlisting}
}

CEP engines like Siddhi allow composition of queries as a \emph{Directed Acyclic Graph (DAG)}, where the output stream from one query feeds in as the input stream for one or more queries. Such dataflows help utility managers design meaningful event analytics from modular queries. Many such event analytics will be active at a time. Some will be persistent to drive dashboards and alerts, while others are exploratory or personalized for individual buildings or analysts. These event analytic dataflows will be registered/deregistered with our IoT platform for deployment onto available resources for execution on the relevant streams. Reducing the end-to-end latency for a given dataflow is one of the Quality of Service (QoS) metrics.

Besides $10-100's$ of such Pi's active in a typical campus setup, 
we also access pay-as-you-go Cloud VMs at Microsoft's Azure Data Center in Singapore where the backend services run. The Pi's are powered by solar panels and rechargeable batteries. While the Pi's can be used for performing CEP queries within the compute, memory and network constraints of each device, we should also ensure that the battery of these gateways must not fully drain, and cause even the basic sensing capability to be lost.

\section{Optimization Problem for Query Placement}
\label{sec:problem}
Here, we formally define the problem of scheduling CEP queries present in analytic dataflows that dynamically arrive and depart, onto edge and Cloud resources. This has the goal of reducing the overall latency of the running dataflows, while meeting compute, network and energy constraints. In \S~\ref{sec:prelim:single}, we introduce and reuse notations from our earlier work on static placement~\cite{ghosh:tcps:2017}, and use these in \S~\ref{sec:prelim:multi} onward to formalize the dynamic variant of the optimization problem.

\subsection{Preliminaries: Single Analytic Dataflow}\label{sec:prelim:single}
The \emph{analytic dataflow} is represented as a \textit{Directed Acyclic Graph (DAG)} of vertices and edges: \( \mathcal{G} = \langle \mathcal {V},\mathcal {E} \rangle \), where $\mathcal{V}=\{v_i\}$ is the set of \textit{CEP queries} that are the vertices of the DAG, and $\mathcal{E}=\langle v_i, v_j \rangle$ is the set of \textit{event streams} that pass the output from $v_i$ to the input of $v_j$~\cite{ghosh:tcps:2017}. Source queries ($\mathcal{V}^{SRC}_{i}$) serve as a dummy input to the DAG, representing the source sensor stream and do not have any predecessor queries, while sink queries ($\mathcal{V}^{SNK}_{i}$) are the final output from the analytic and do not have successors.
We assume that the output events of a query are \emph{duplicated} across all outgoing edges from that vertex and the inputs for a query from multiple incoming edges are \emph{interleaved}. 
A \emph{path} $p_i \in \mathcal{P}$ of length $n$ in the graph $\mathcal{G}$ is a connected sequence of $n$ edges with $n+1$ vertices, starting at a source query and ending at a sink query. $\mathcal{P}$ is the set of all paths in the DAG.

\textit{Stream rate} is the number of events passing per unit time on a stream~\cite{ghosh:tcps:2017}. The \textit{input stream rate}, $\Omega^{in}$, to a dataflow is the sum of the output stream rates from all source queries in the DAG, while the \textit{output stream rate}, $\Omega^{out}$, for the dataflow is the sum of output stream rates from the sink queries. A \textit{selectivity} function $\sigma(v_i)$ gives the average number of output events expected for each input event processed by a query. 
The \textit{input stream rate}, $\omega^{in}_{i}$, for a vertex $v_i$ is the sum of the stream rates on all its incoming edges and its \textit{output stream rate} is the product of its incoming stream rate and its selectivity, $\omega^{out}_{i} = \omega^{in}_{i} \times \sigma(v_i)$. We can then recursively compute the input and output rates for downstream queries $v_j$, and the output rate for the entire DAG. For simplicity, if the output stream rate for all source queries $v_k \in \mathcal{V}^{SRC}$ is uniform, we have $\omega^{out}_k = \frac{\Omega^{in}}{|\mathcal{V}^{SRC}|}$. The \textit{selectivity for the whole dataflow} is $\sigma(\mathcal{G})=\frac{\Omega^{out}}{\Omega^{in}}$.

We consider two classes of computing \emph{resources} -- $\mathbb{R}_E$ for edge devices and $\mathbb{R}_C$ for Cloud VMs, with the set of all computing resources available in the IoT fabric given by $\mathbb{R} = \{r_k\} = \mathbb{R}_E \cup \mathbb{R}_C$ and $\mathbb{R}_E \cap \mathbb{R}_C=\varnothing$~\cite{ghosh:tcps:2017}. A CEP query in a dataflow executes on a single resource $r_k$, and a \textit{resource mapping function} indicates this, $\mu: \mathcal{V} \rightarrow \mathbb{R}$. 

\textit{Compute latency} (or latency) denoted by $\lambda_{i}^{k}$ is the time taken to process one event by a query $v_i$ on an exclusive resource $r_k$. This will depend on both the query type as well as the resource type.  If the \textit{size of an event} that is emitted by the query on its outgoing edge(s) is $\delta_{i}$, and the \emph{network latency} and \emph{network bandwidth} between two resources $r_m$ and $r_n$ is given by $l_{m,n}$ and $\beta_{m,n}$, respectively, then the \emph{end-to-end latency} along a path $p \in \mathcal{P}$ for a given resource mapping $\mu$ for the DAG is defined as,
\[L_p =  \sum \limits_{\substack{ \langle v_i,v_j \rangle \in p\\(v_i, r_m) \in \mu\\(v_j, r_n) \in \mu}} \Big(\lambda_{i}^{m} + \big(l_{m,n} + \frac{\delta_{i}}{\beta_{m,n}}\big) \Big)\] 
The maximum over the end-to-end latency along all paths $p \in \mathcal{P}$ gives us the \emph{makespan} for the DAG for the given mapping, with the maximum path called the \emph{critical path}.
\[L_{\mathcal{G}} = \max_{\forall p ~\in \mathcal{P}}(L_p)\]

\subsection{Preliminaries: Dynamic Analytic Dataflows}\label{sec:prelim:multi}

We extend the single static dataflow above to a dynamic situation where analytics arrive and depart the system over time. We represent the \emph{set of active dataflows} at logical time $t$ as $\mathbb{G}^t = \{ \mathcal{G}_0, \mathcal{G}_1, ..., \mathcal{G}_n \}$, where $\mathcal{G}_i = \langle \mathcal{V}_i, \mathcal{E}_i \rangle$ is one active dataflow with CEP queries $v_{i,k} \in \mathcal{V}_i$. 
They have corresponding \emph{input} and \emph{output stream rates} of $\Omega^{in}_{i}$ and $\Omega^{out}_{i}$, respectively. Their selectivities are denoted as $\sigma(\mathbb{G}^t) = \{\sigma(\mathcal{G}_0), \sigma(\mathcal{G}_1), ..., \sigma(\mathcal{G}_n)\}$, where 
$\sigma(\mathcal{G}_i) = \frac{\Omega^{out}_{i}}{\Omega^{in}_{i}}$.
When these dataflows are mapped to a set of edge and Cloud resources $\mathbb{R}$, their \emph{mapping functions set} at time $t$ is represented as
$\mathbb{M}^t = \{\mu_0^t, \mu_1^t, ..., \mu_n^t\}$,  where $\mu_i: \mathcal{V}_i \rightarrow \mathbb{R}$.


The system of analytic dataflows deployed on the set of IoT resources at time $t$ may undergo a change at each subsequent time interval. We define a \emph{control interval}, $\theta$, as the time period at which dataflows can arrive or depart. This is the interval at which (re)mapping decisions should be made.  
At each control interval, we consider three possible \emph{activities} by the users: a new DAG \textit{arrives}, an existing DAG \textit{departs}, or \emph{no change} happens, and we take appropriate scheduling \emph{actions}. For convenience, we increment the logical time $t$ in units of $\theta$ so that $t+1$ indicates the next control interval. Here, we assume that at each control interval, only a single dataflow may arrive or depart, and the dataflows' input rates are stable. But these can be generalized in future.

As the analytic dataflows arrive and depart, 
it is necessary to deploy new queries onto available resources, or stop old queries and release resources. Additionally, remapping of queries of other active dataflows may ensure that their performances are not affected. 
%
%
Let at time $t$, the mapping functions for the active dataflows $\mathbb{G}^t$ be the set $\mathbb{M}^{t}$. If a new DAG arrives or leaves, the set $\mathbb{M}^{(t+1)}$ will have a new mapping added, an old mapping removed, and/or reconfigurations of mappings of dataflows that stay active. The time taken to find the new mappings for the active DAGs at time $t$ is the \emph{schedule planning time} $\phi_t$.

A \emph{reconfiguration} of the mapping for a dataflow that continues to be active after a control interval will cause a change in its prior mapping of queries to resources. 
We define a binary function $\rho_{t}(v_{k,i})$ for a vertex $v_{k,i}\in\mathcal{V}_k$ in DAG $\mathcal{G}_k$ at time $t$ to capture the occurrence of a reconfiguration, 

\[ \rho_{t}(v_{k,i}) = 
\begin{cases} 
1 &  \langle v_{k,i}, r_p \rangle \in \mu^t_k,~~\langle v_{k,i}, r_q \rangle \in \mu^{t+1}_k,~~r_p \neq r_q\\
0 &  \text{\emph{otherwise}}
\end{cases}
\]
\emph{Migration time} is a constant time $\eta$ taken for moving a CEP query $v_{k,i}$ from resource $r_p$ to $r_q$ upon reconfiguration. The total number of migrations at control interval $t$ is given by, \[\overline{\rho_t} = \sum \limits_{\forall v_{k,i} \in \mathcal{V}_{i},~\mathcal{V}_{i} \in \mathbb{V}^{t}} \rho_t(v_{k,i})\]

When a query of a DAG $\mathcal{G}_k$ is migrated, the input stream for the query is buffered for later downstream processing after the schedule has been enacted. Given an input stream rate of $\omega^{in}_{k,i}$ for a query $v_{k,i}$, the number events that will be \emph{buffered} in a queue during a control interval action is,
\[q_{k,i} = \omega^{in}_{k,i} \times \eta ~~\text{such that}~~ v_{k,i} \in \mathcal{V}_k, \rho_{t} (v_{k,i}) = 1\]
%
%
After the schedule has been enacted, these events buffered during the migration have to be processed along with events that continue to arrive on the input streams to the DAG. The amount of time required for a query $v_{k,i}$ to process and drain all buffered events, and catch-up to a stable rate is called the \emph{stabilization time} $\psi_{k,i}$. If the latency for processing an event for a vertex $v_{k,i}$ mapped on a new resource $r_p$ is $\lambda^p_{k,i}$, stabilization time for that vertex can be calculated as,
\[\psi_{k,i} = \frac{q_{k,i}}{\frac{1}{\lambda^p_{k,i}} - \omega^{in}_{k,i}} \]
which is the buffered queue size divided by the difference between the supported input rate and current input rate.

The total stabilization time $\overline{\psi}$ for a collection of dataflows $\mathbb{G}$ after reconfigurations at time $t$ is given by,
\[\overline{\psi_t} = \max \limits_{\substack{\forall k,~\forall v_{k,i} \in \mathcal{V}_k,~\rho_{t} (v_{k,i}) = 1}}\big( \psi_{k,i} \big) \]

\subsection{Query Placement Constraints}
\label{sec:problem:dyn:cons}
Based on the earlier motivating scenario, we define several constraints to be met when deciding the placement of queries to edge and Cloud resources. These are similar to our earlier work on static query placement~\cite{ghosh:tcps:2017}, but modified for the dynamic scenario that we now consider.
\begin{cons}\label{cons:dyn:1}
	Source queries in all DAGs must be placed on edges, while the sink queries must be placed on the Cloud.
	\begin{align*}
	\forall \mu^{t}_{k},~ \langle v_{k,i}, r_p \rangle \in \mu^{t}_{k} ~~~&|&~~~ v_{k,i} \in \mathcal{V}^{SRC}_{k} &\implies r_p \in \mathbb{R}_E \\
	&|&~~~ v_{k,i} \in \mathcal{V}^{SNK}_{k} &\implies r_p \in \mathbb{R}_C
	\end{align*}
\end{cons}
\noindent Event analytics operate on streams sourced from the edge but often have costly control decisions occurring in the Cloud. Hence, the dataflow should consume events from the edge and deliver results to the Cloud. Thus, the decision making  responsiveness will depend on the end-to-end latency across the edge and Cloud network. This forces source queries to be co-located on edge devices that generate the input stream(s), and sink queries to be placed on VMs.

\begin{cons}\label{cons:dyn:2}
	Given an input rate $\omega^{in}_{k,i}$ to query $v_{k,i}$ of DAG $\mathcal{G}_k$, the query must not overwhelm the compute capacity if \emph{exclusively} mapped to a resource $r_p$.
	\[\omega^{in}_{k,i} < \frac{1}{\lambda^p_{k,i}} ~~\forall v_{k,i} \in \mathcal{V}_k\]
	If \emph{multiple queries} from one or more DAGs run on the same resource $r_p$, then the input rate $\omega^{in}_{k,i}$ on a vertex $v_{k,i}$ that the resource can handle is limited by:
\[\omega^{in}_{k,i} ~~<~~ \frac{1}{\sum \limits_{\mu^{t}_k}~\sum \limits_{\substack{(v_{k,j}, r_p) \in \mu^{t}_k \\v_{k,j} \notin \mathcal{V}^{SRC}_k}}\big(\lambda^p_{k,j}\big)} \Big(1 + \pi(m)\Big)\]
\[m = \sum \limits_{(v_{k,j}, r_p) \in \mu^{t}_k} |v_{k,j}|, \forall~~ v_{k,i} \in \mathcal{V}_k,~ v_{k,i} \notin \mathcal{V}^{SRC}_k\]

\end{cons}

\noindent The maximum input rate that a resource $r_p$ can handle when exclusively running a query $v_{k,i}$ is the inverse of its latency $\frac{1}{\lambda^p_{k,i}}$, and for multiple queries it is the inverse of the sum of their latencies $\frac{1}{\sum\lambda^p_{k,i}}$. However, there is likely to be additional overheads in the latter case due to interference between concurrent queries. If $m$ queries are running on a resource $r_p$, let $\pi(m)$ denote the \emph{parallelism overhead}, which is obtained empirically. Hence, we should only place a query on a resource if it will not receive an input rate greater than this upper-bound throughput.

\vspace{0.2cm}
Lastly, edge resources on the field are often powered by rechargeable batteries that are connected to renewables like solar panel to reduce their maintenance. Let $C_p$, in $mAh$, be the \emph{power capacity} of a fully-charged battery for the edge device $r_p\in \mathbb{R}_E$. Let the \emph{base load} (instantaneous current) drawn by the edge device when no queries are running be $\kappa^p$, in $mA$. Let $\epsilon_{k,i}^{p}$ be the \emph{incremental power}, in $mAh$ beyond $\kappa^p$, drawn on the edge resource by a query $v_{k,i}$ to process a single input event. Let the \emph{recharge interval} to fully recharge this battery be $\tau_p$, in $seconds$, say through solar generation or by replacing the battery.
\begin{cons}\label{cons:dyn:3}
	The queries running on a edge device $r_p$ should not fully drain its battery capacity within the recharge time interval $\tau_p$.
	
	\[ \tau_p \times \Big( \kappa^p + \sum \limits_{\mu^{t}_k}~ \sum \limits_{\substack{(v_{k,i}, r_p) \in \mu^{t}_k\\v_{k,i} \notin \mathcal{V}^{SRC}_k\\r_p \notin \mathbb{R}_C}}(\omega^{in}_{k,i} \times \epsilon_{k,i}^{p}) \Big) ~~\leq~~ C_p\]
\end{cons}
 
\noindent The first term is the base load that drains the edge resource, even when inactive, during the recharge interval. The second term is the incremental power for processing events by all queries mapped to that edge at their respective input rates.

%

\subsection{Optimization Problem}
Given a set of DAGs $\mathbb{G}^t = \{\mathcal{G}_{0}, \mathcal{G}_{1}, ...,  \mathcal{G}_{n}\}$ which have been scheduled at time $t$ on a set of Edge and Cloud resources $\mathbb{R}$, denoted by a set of resource mappings $\mathbb{M}^{t} = \{\mu_{0}^{t}, \mu_{1}^{t}, ..., \mu_{n}^{t}\}$, without violating the above three constraints. When a DAG arrives or leaves the system at time $t+1$, the primary objective is to find a new mappings set $\mathbb{M}^{t+1}$ that meets Constraints~\ref{cons:dyn:1}, \ref{cons:dyn:2} and \ref{cons:dyn:3}, while \emph{mini\-mi\-zing the sum of the makespans} for all the DAGs, $\mathbb{G}^{t+1}$, given by,

\[ \widehat{L}_{\mathbb{G}^{t+1}} = \sum \limits_{\mathcal{G}_i \in \mathbb{G}^{t+1}} ~ \min_{\forall \mu^{t+1}_i : \mathcal{V}_i \rightarrow \mathbb{R}} \big(L_{\mathcal{G}_i} \big) \] 

\noindent Secondary objectives to this optimization problem are to minimize the \emph{schedule planning time} $\phi_{t+1}$ at the control interval $t+1$, the \emph{total number of migrations} performed, $\overline{\rho_{t+1}}$, and the \emph{total stabilization time} $\overline{\psi_{t+1}}$.

\section{Adaptive Placement Strategies}
\label{sec:solution}

The solution to the above optimization problem is NP-complete as optimal DAG scheduling in general is NP-complete~\cite{Kwok99}. As DAGs arrive and depart, placing them on or removing them from existing resources will affect the latency time and the constraints of the other DAGs that are collocated on the same resource(s). As a result, an optimal placement that minimizes the sum of makespans of all active DAGs may require all the queries for all active DAGs to be rescheduled. The time complexity for a brute force solution to this at a given control interval $t$ is exponential in terms of the number of resources, $\mathcal{O}\big((|\mathbb{V}^t|+|\mathbb{E}^t|)\times|\mathbb{R}|^{n}\big)$. This take days to solve optimally even for a single dataflow with $14$ queries on, say, 50 resources~\cite{ghosh:tcps:2017}. Hence we explore heuristics that offer a reasonable quality solution to minimize the makespan sum while guaranteeing that constraints are met.

In our proposed approach, we perform several actions at each control interval. When a dataflow \emph{arrives}, we need to determine on which available resources to place its queries on while reducing its end-to-end latency. We should meet its constraints, and also ensure that the constraints of existing dataflows on those resources continue to be met. We propose a novel \emph{dataflow scheduling heuristic} for performing this in \S~\ref{sec:solution:heu}, and further extend a prior GA-based meta-heuristics to this dynamic scenario in \S~\ref{sec:solution:ga}. When a dataflow \emph{departs}, we need to stop its CEP queries and reclaim its resources. This will not violate the constraints of existing dataflows. 

Once the dataflow in question has been mapped/unmapped within the constraints, we next check if the sum of dataflow makespans can be improved by reconfiguring the dataflow queries through migrations, while also minimizing the number of migrations which will affect the stabilization time. We propose several \emph{rebalancing} strategies for these selective migrations in \S~\ref{sec:solution:rebal}. Both the dataflow scheduling heuristic and the rebalance strategies contribute to the schedule planning time, which must be reduced so that these online algorithms complete within each control interval.




\subsection{Topological Set Ordering (TopSet) Heuristic }\label{sec:solution:heu}
The makespan for each event analytic dataflow that arrives is determined by its critical path. \emph{Topological sorting} is frequently used for DAG scheduling~\cite{Kwok99}, where the queries in the DAG are traversed in a BFS order starting from the source tasks and scheduled on the most suitable available resource, in that order. Others use a rank based approach that assigns a priority for each query in the DAG based on its presence in the critical path from the source to the sink tasks~\cite{zhao2006scheduling}. However, these are designed for batch workflows, rather than streaming dataflows, and for scheduling a single workflow rather than dynamic dataflows. We adapt these techniques for our heuristic by extending the topological DAG ordering with a local ranking at each level.

When a DAG $\mathcal{G}_k$ arrives, our \textbf{TopSet} heuristic traverses it in topological order that ensures that all parent (upstream) queries are visited before their child queries. Further, we rank the children of a parent so that they are visited in decreasing order of query latency. 
Such an ordering is obtained by finding the \emph{topological set} ordering. When performing the multi-source BFS traversal, instead of appending a child to the topological list, we merge all children at the same depth for a parent into one set. This traversal will return a list of query sets, with each set having sibling queries and with the previous set in the list referring to the parent set. 

Formally, a topological list of sets $\mathbb{S}_k = [ S_i ]$ for the traversal of DAG, $\mathcal{G}_k=(\mathcal{V}_k,\mathcal{E}_k)$, is defined as a recurrence,
	\begin{align*}
	S_0 &= \{ v \in \mathcal{V}_k \mid \forall u \in \mathcal{V}_k, \langle u,v \rangle \notin \mathcal{E}_k \} \\
	S_{i+1} &= \{v \in \mathcal{V}_k \mid u \in \bigcup_{j=0}^i S_j ,~~\forall u \in \mathcal{V}_k,\langle u,v \rangle \in \mathcal{E}_k \}
	\end{align*}
\noindent where $S_0$ is the source set containing all source queries.

\ysnoted{it is not clear in what order you visit within a set}
Given $\mathbb{S}_k$ for the incoming DAG, we \emph{visit} each set in the list sequentially, and within each set visit the queries in decreasing order of critical latency to consider it for placement on the available resources. The capacity of resources $\mathbb{R}$ is a function of the concurrent queries from existing dataflows running on each. When a query $v_{k,i}$ is visited, we evaluate its quality on each resource $r_p \in \mathbb{R}$ by calculating the critical path length from the source queries to this query.
\ysnoted{I thought the latency does not change when we place multiple queries on a resource, only the input rate supported changes?}
		\[L_i = \max \limits_{\substack{\forall v_{k,j} \in parents(v_{k,i})\\(v_{k,j}, r_q) \in \mu_k}} \Big(\lambda_{k,i}^p + \big(l_{q,p} + \frac{\delta_i}{\beta_{q,p}} \big)\Big)\] 
When we place the query $v_{k,i}$ on a resource $r_p$ which already has other upstream queries from the dataflow placed in it, there will be an impact on the latencies of the previous queries due to interference. We optionally assign a \emph{penalty} on the current query's latency for that resource. 
This penalty is the sum of increase in latency of the critical path lengths $L_j$ for all queries $v_{k,j}$ placed on the resource $r_p$, relative to their previously estimated critical path length. We term this variant as \textbf{TopSet/P}. 

We consider a resource as a valid mapping for a visited query only if it does not violate the three constraints, either for this query or for prior queries placed on it. Among the valid resources, we select the one with smallest critical path latency for mapping this query, and add it to the mapping function for the DAG, $\mu_k$. We expect a query to be more likely to be placed on an edge resource till the latency/capacity of the edge surpass the network latency from the edge to the Cloud. Once an edge violates the constraints, queries are likely to move to Cloud resources.

TopSet only schedules queries of arriving DAG on the available resource. It does not migrate queries in existing dataflows. It also does not handle DAG removals explicitly, and we just remove the queries of the departing DAG from the resources they were placed on. This may improve the performance of remaining DAGs, but this improvement may be sub-optimal. Later, we discuss \emph{rebalancing} strategies to globally reconfigure queries across all active dataflows.

\subsection{Extensions to GA Meta-heuristics}\label{sec:solution:ga}
In our prior work, we reduced the single dataflow scheduling problem to a \emph{Genetic Algorithm (GA)} formulation~\cite{ghosh:tcps:2017}. That approach models a \emph{chromosome} as the mapping function $\mu$, with a length $n$ that matches the number of queries in the dataflow and each \emph{basepair} having a value from $[0, |\mathbb{R}-1|]$. The GA algorithm generates a random \emph{population} of chromosomes, and uses \emph{crossovers} and \emph{mutations} to create a new \emph{generation} of population~\cite{Michalewicz96}. In each population, we penalize chromosomes that violate any constraint. We keep track of the chromosome which gives the best makespan among all valid solutions seen. The algorithm terminates after a fixed number of generations, or if no improvement is seen in the past $50\%$ of generations.
 
We propose two extensions to this GA algorithm to support the current problem of having multiple dataflows arriving and departing dynamically from the system.

\para{GA-Incremental (GAI)}
When a new DAG arrives, we use GA to schedule its queries on the available resources, after reducing their capacities based on the previous queries running on them. We obtain the latency $\lambda_k$, energy $\epsilon_k$ and throughput rate $\omega^{in}_k$ supported by the resources from earlier deployments to drive this GA placement. We run GA on the DAG with these updated resource capacities. 
If the GA converges to a valid placement, we deploy the DAG on the resources returned by the chromosome mapping function. If the GA does not converge due to constraint violations, we discard the DAG and report an error. The resource mappings for prior dataflows are retained, and only queries in the arriving DAG are incrementally scheduled. As for TopSet, GAI does not handle DAG departures other than reclaiming those resource capacities for future DAG arrivals.

This approach is simple by reusing an earlier algorithm for a static scenario, and requires no reconfiguration of existing dataflows. However, GA while faster than a brute-force approach still takes longer to converge than our TopSet heuristic which has a bounded time. GA cannot guarantee an optimal solution either, as with similar meta-heuristics~\cite{Michalewicz96}. So it is ill-suited for online scheduling of dynamic DAGs.

%

\para{GA-Global (GAG)}
Another variant of GA addresses the potential sub-optimal global state of dataflows after a single dataflow has been placed or an existing dataflow unmapped by TopSet or GAI. When a DAG arrives or departs the system at time $t$, the set of active DAGs $\mathbb{G}^t$ changes. GAG considers this entire set of dataflows for \emph{ab initio} placement, irrespective of their prior placement on resources. We translate the DAGs in $\mathbb{G}$ into a single \emph{global DAG} by connecting all their source queries to a dummy source and all their sinks to a dummy sink. GA is then run on this global DAG to place \emph{all queries} in all active DAGs on the set of all resources whose \emph{full capacities} are available.


This is expected to give a better solution for the optimization problem than GAI 
since all DAGs (and not the most recently arriving DAG) 
are considered for placement. However, such an approach results in high migration costs since the placement for all queries, both new and previously placed, will change at each control interval. 
This will also have a larger schedule planning time relative to GAI. 

\subsection{Rebalancing of Placements}\label{sec:solution:rebal}
It is common for DAG scheduling heuristics to start with an approximate placement and then perform a constrained search to incrementally improve this solution~\cite{wu2001efficient}. We propose \emph{rebalancing strategies} that start from a valid schedule from one of the above heuristics, and then improves upon their solutions to reduce the overall makespan of dataflows. These specifically look at the queries on the critical path of different dataflows and try to migrate these queries to better resources, provided that no constraints are violated.

\para{Vertex Rebalancing} In this strategy, the query having the highest latency along the critical path of a DAG is migrated to a resource which has a higher compute capacity, thereby reducing the critical path length of the DAG. When a vertex rebalance is performed, the DAGs currently present in the collection $\mathbb{G}^t$ at time $t$ are sorted in decreasing order of their critical path latency. Each DAG is visited in the sorted order obtained, and the query with highest latency along its critical path is chosen for rebalance. This query is moved to a resource which reduces the objective value of the optimization problem. Such a relaxation of the costliest query in the critical path for each active DAGs results in a maximum of $|\mathbb{G}^t|$ migrations.

\para{Edge Rebalance} Besides query latency due to the compute capacity of a resource, network latency forms the other major factor in the critical path. This will also be affected by the network latencies between different resources deployed in a wide area network. Edge rebalance also sorts the DAGs in decreasing order of critical path latencies. In each path, it identifies the edge with the highest network cost (latency and bandwidth) for rebalance. We then test if moving the upstream query of this critical edge to the same resource as the downstream query will improve the makespan, or if moving the downstream query to the upstream resource will help. 
We pick the operation that offers the better improvement. A maximum of $|\mathbb{G}^t|$ migrations may be performed. Code Repository is available online \footnote{Code is available at \url{https://github.com/dream-lab/ec-sched}}

\section{Results}
\label{sec:results}
\begin{figure}[t]
	\centering
	\includegraphics[width=0.7\columnwidth]{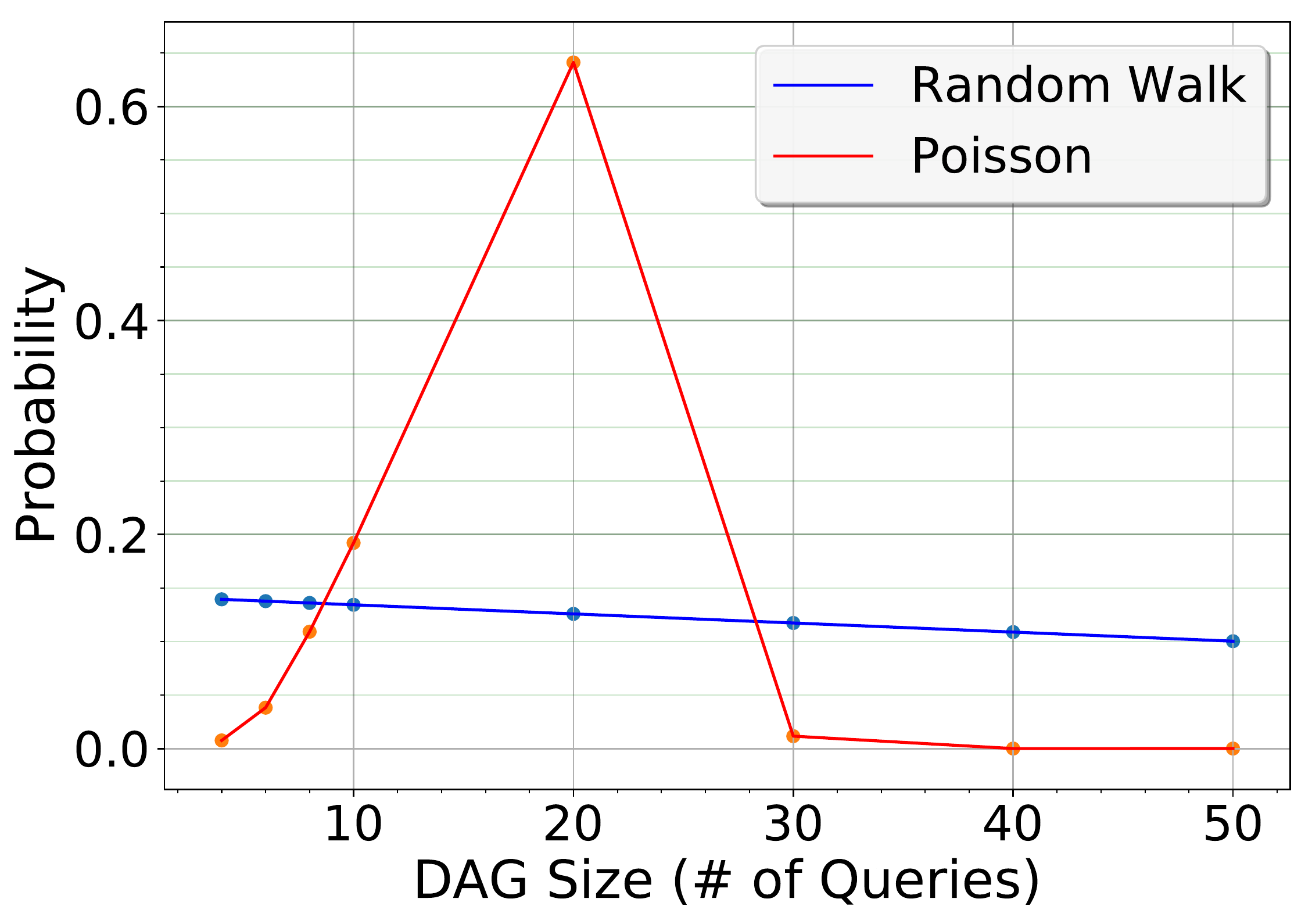}
	\caption{DAG size distributions for workload generation}
	\label{fig:prob}
\vspace{-0.15in}
\end{figure}
\begin{figure*}[t!]
	\centering
	\subfloat[Random Walk, $\widehat{U} = 2.0 \pm 0$]{
		\includegraphics[width=0.23\textwidth]{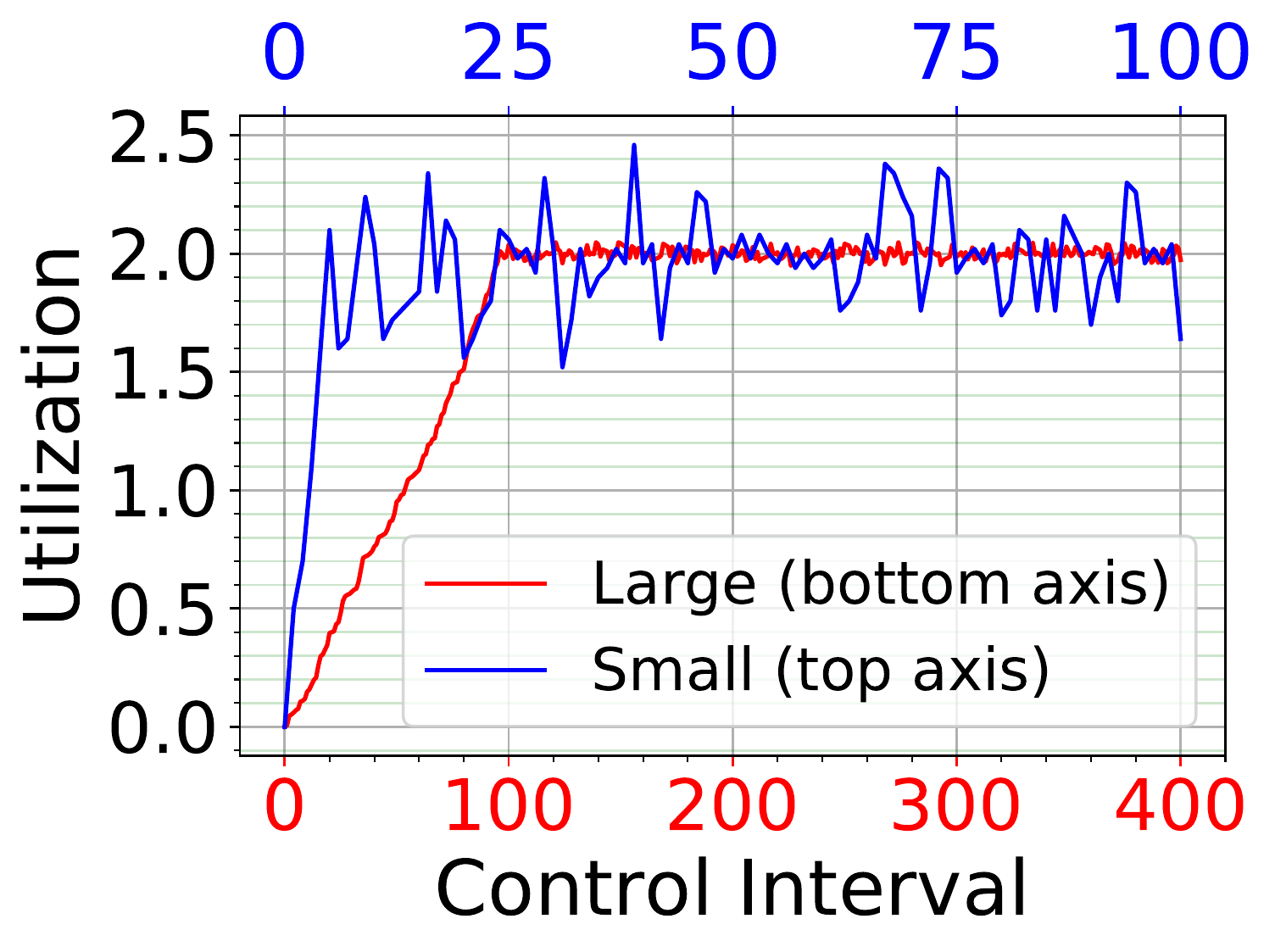}
		\label{fig:workload1}
	}
	\subfloat[Random Walk, $\widehat{U} = 2.0 \pm 0.5$]{
		\includegraphics[width=0.23\textwidth]{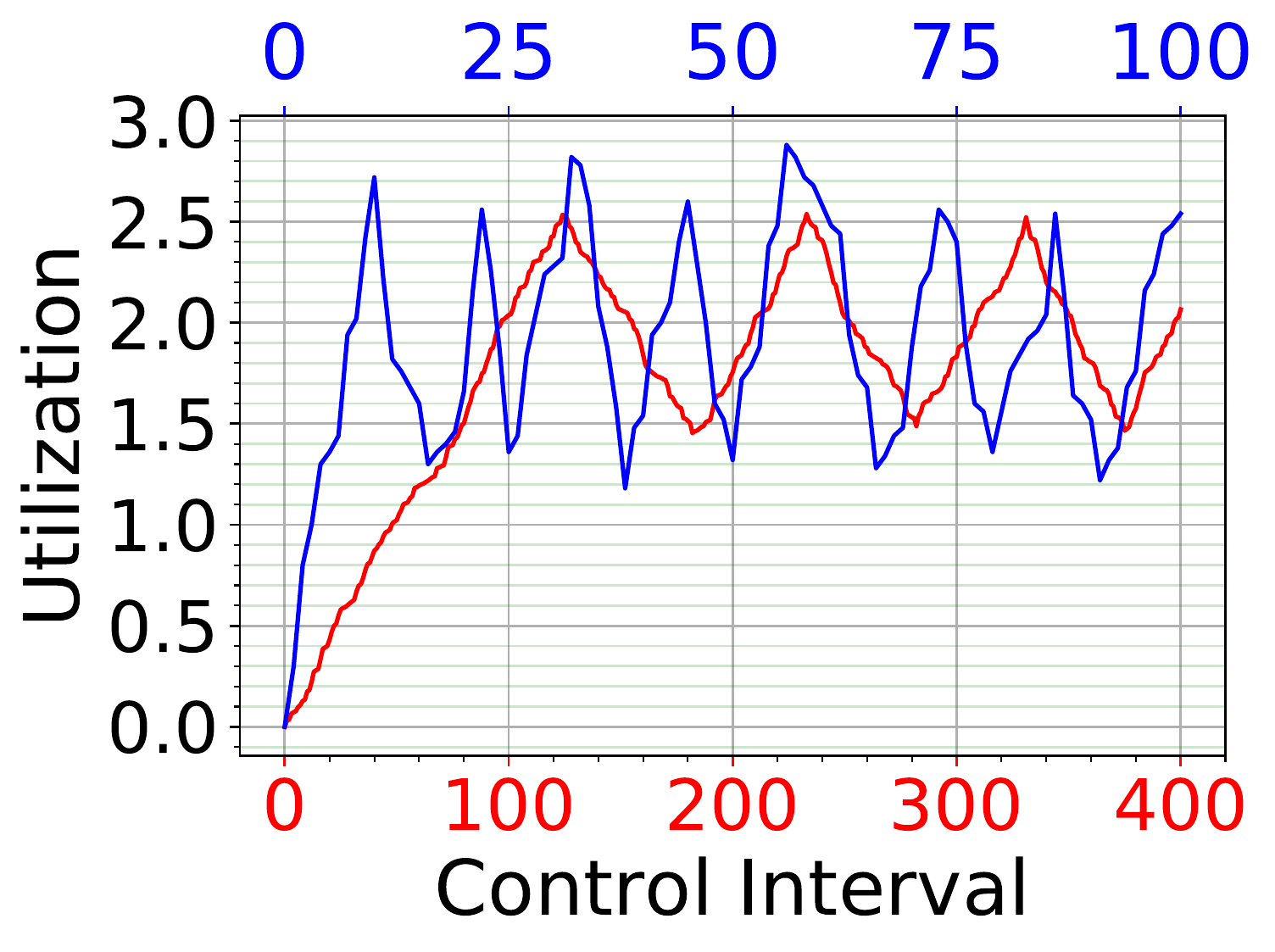}
		\label{fig:workload2}
	}
	\subfloat[Random Walk, $\widehat{U} = 2.0 \pm 1.0$]{
		\includegraphics[width=0.23\textwidth]{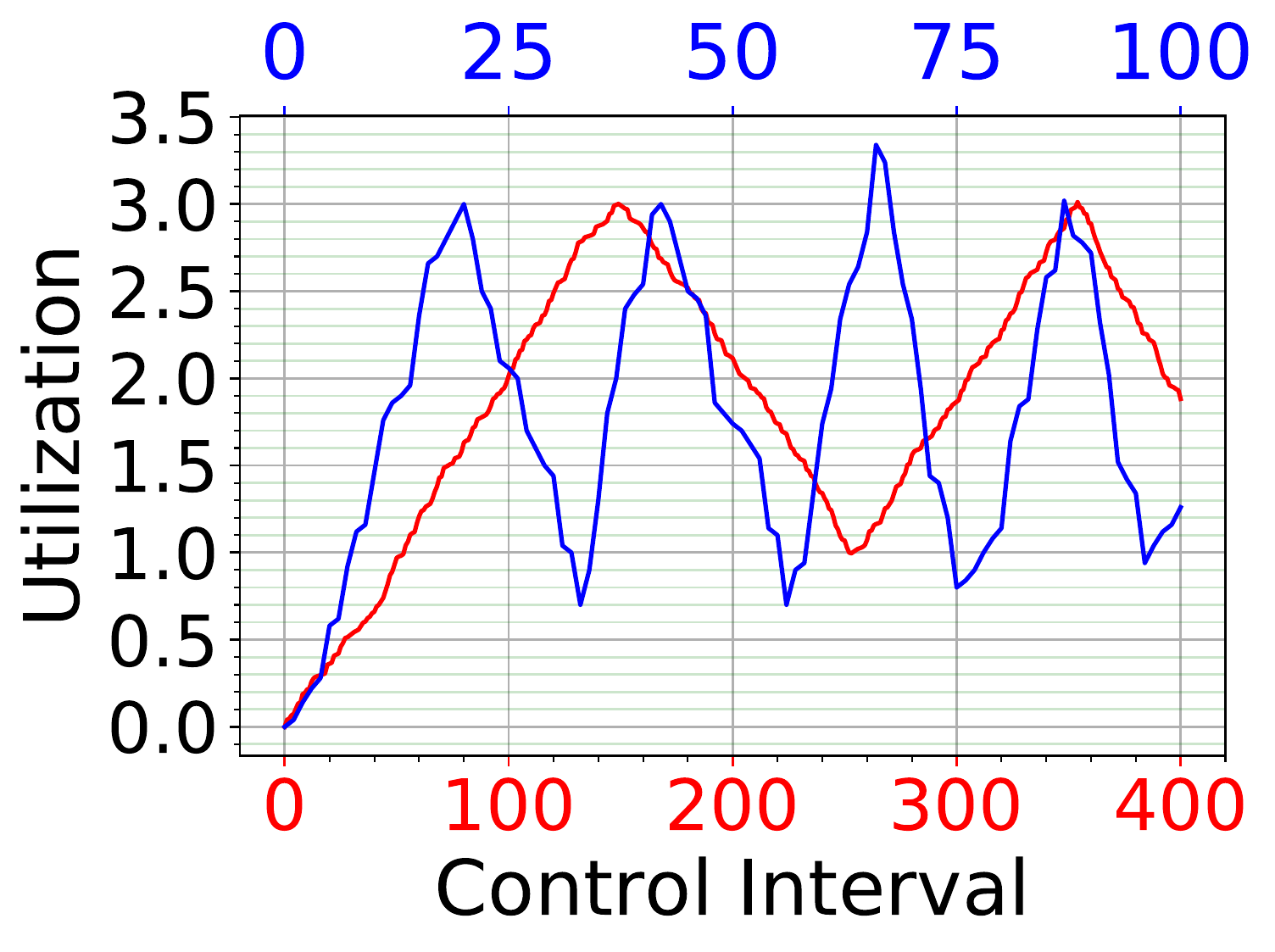}
		\label{fig:workload3}
	}
	\subfloat[Poisson, $\lambda=12$]{
		\includegraphics[width=0.23\textwidth]{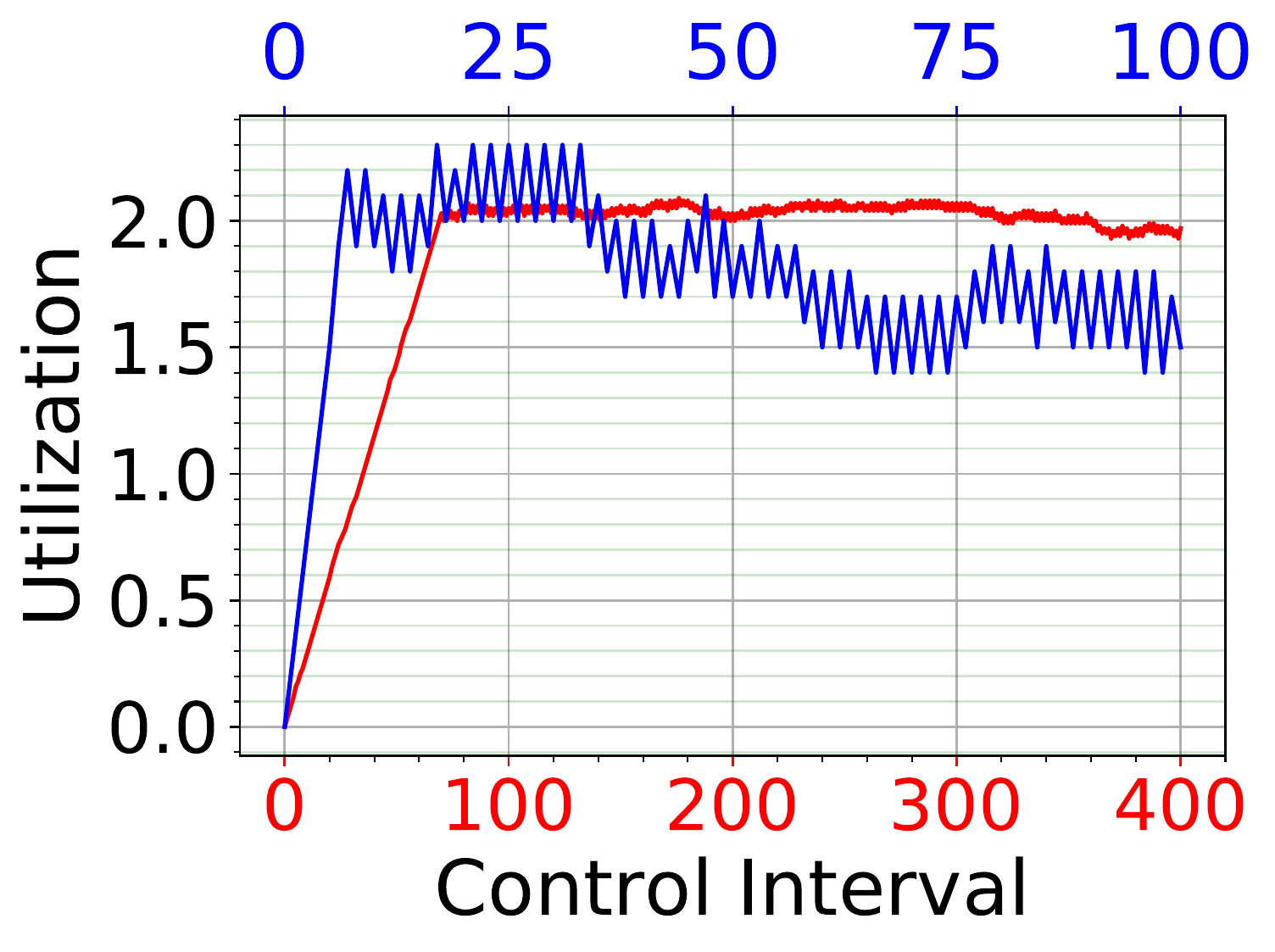}
		\label{fig:workload4}
	}
\vspace{-0.05in}
	\caption{$\mathcal{U}^t$ across time for the workloads. \emph{Small} has $100$ devices/$100$ intervals and \emph{Large} has $1,000$ devices/$400$ intervals.}
	\label{fig:workload}
	\vspace{-0.15in}
\end{figure*}

We perform a realistic simulation study to evaluate the dynamic dataflow scheduling strategies. These use detailed traces from real-world micro and application benchmarks on the Campus IoT deployment, on edge and Cloud resources. 

\para{System Setup}
We use \emph{Raspberry Pi 2 Model B v1.1} as our edge devices, with a $900MHz$ 4-core ARM A7 CPU, $1~GB$ RAM and $100~Mbps$ NIC. A \emph{Microsoft Azure D2 VM} in Southeast Asia data center serves as our Cloud resource, with a $2.2Ghz$ 2-core Intel Xeon E5 CPU, $7~GB$ RAM and Gigabit NIC. Both resource types run Linux and WSO2's Siddhi CEP engine within a JRE~\cite{siddhi11}. The edge devices are connected by Gigabit Ethernet campus network, and access the Azure data center through the public Internet. In our IoT deployment for the simulation, we used two resource setups: \emph{small}, one with $96$ Pi devices and $4$ Azure VMs for a total of $100$ resources, and \emph{large}, with $960$ Pi devices and $40$ Azure VMs for a total of $1,000$ resources. This captures a campus and a township scale IoT deployments, and bounds the operational costs for the pay-as-you-go Cloud VMs.

We identify $21$ different CEP query types that span filter, sequence, pattern, sliding window aggregate and batch window aggregate, and with different configurations, such as sequence length, window length and selectivity. These queries are individually micro-benchmarked on the Pi and the Azure VM to measure various coefficients used in \S~\ref{sec:problem}, such as their peak compute throughput, parallelism overhead, base load and incremental power consumption, etc. We also make detailed network latency and bandwidth measurements between pairs of edge devices, and the edge and Cloud VM. During the simulation runs, we sample values from these distributions to capture the variability in performance of even identical resources when operating in the field. For brevity, we refer the readers to the detailed benchmark measurements from earlier~\cite{ghosh:tcps:2017}.

\para{Simulation Workloads}
We generate $39$ different event analytic dataflows using the \emph{Random Task and Resource Graph (RTRG)} utility~\cite{RTRG}, with $4-50$ vertices each and a fan-out of up to $5$ edges. We uniformly sample and assign one of the benchmarked CEP queries to each vertex. We use a constant cumulative input rate of $100~e/sec$ for each dataflow, though the input rate per source query can range from $25-100~e/sec$. We ensure coverage of query types ($21$), selectivities ($0.01 - 458$), output rates ($1-11,457~e/sec$), source queries ($1-4$), sink queries ($1-3$), etc. and make sure there is a feasible valid placement for the dataflow on the available resources~\cite{ghosh:tcps:2017}.

We simulate the dynamic arrival and departure of DAGs at each control interval using a Random Walk (RW) and a Poisson distribution~\cite{Arampatzis:2008}. In the \emph{Poisson model}, we \emph{alternate} between adding and removing a DAG after initially adding $16$ and $70$ DAGs for Small and Large resource setups, respectively. In the \emph{RW model}, we use the cumulative resource utilization to decide whether to add or remove a DAG. 
The \emph{utilization} of the resources $\mathbb{R}$ with $\mathcal{G}_i\langle \mathcal{V}_i, \mathcal{E}_i \rangle \in \mathbb{G}^t$ DAGs active at time $t$ is,
\[\mathcal{U}^t = \frac{\sum \limits_{i=1}^{|\mathbb{G}^t|} |\mathcal{V}_i|}{|\mathbb{R}_E|+|\mathbb{R}_C|}\]
We define a \emph{utilization threshold} $\widehat{U} = U \pm u$. We continue adding DAGs at each control interval $t+1$ while $\mathcal{U}^t < (U+u)$, and switch to removing a DAG while $\mathcal{U}^t > (U-u)$. This oscillates DAG addition and removal within an upper and lower bound. We use $U=2$, i.e., an average of $2$ queries per resource, and $3$ values of $u=0, 0.5$ and $1.0$, which allows an average deviation of $\pm u$ queries from $2$.


Once we have decided to add or remove a DAG at a control interval, the size of the DAG is determined from a probability distribution shown in Fig.~\ref{fig:prob}. For RW, the chance that a DAG is selected is inversely proportional to the number of queries it has, while for Poisson, it follows a Poisson distribution with $\lambda=12$ queries, which is close to the median DAG size. 
%
Given the DAG size, we choose a specific DAG to add 
by uniformly selecting from the different variants of this DAG size in the pool of $39$ dataflows; 
or we choose a specific DAG to remove 
by linearly searching through the active DAGs till one with this size is located, and otherwise, 
sample again from the distribution till a size matches an active DAG. We run this simulator for $100$ and $400$ control intervals for the Small and Large resource setups, respectively. Figs.~\ref{fig:workload} show the utilizations for the generated workloads across the control intervals as DAGs are added or removed.
This variation is smoother for the large setup as compared to small since the relative change in utilization due to adding/removing a single DAG is smaller.

\begin{figure*}[t!]
	\centering
	\subfloat{
		\includegraphics[width=0.23\textwidth]{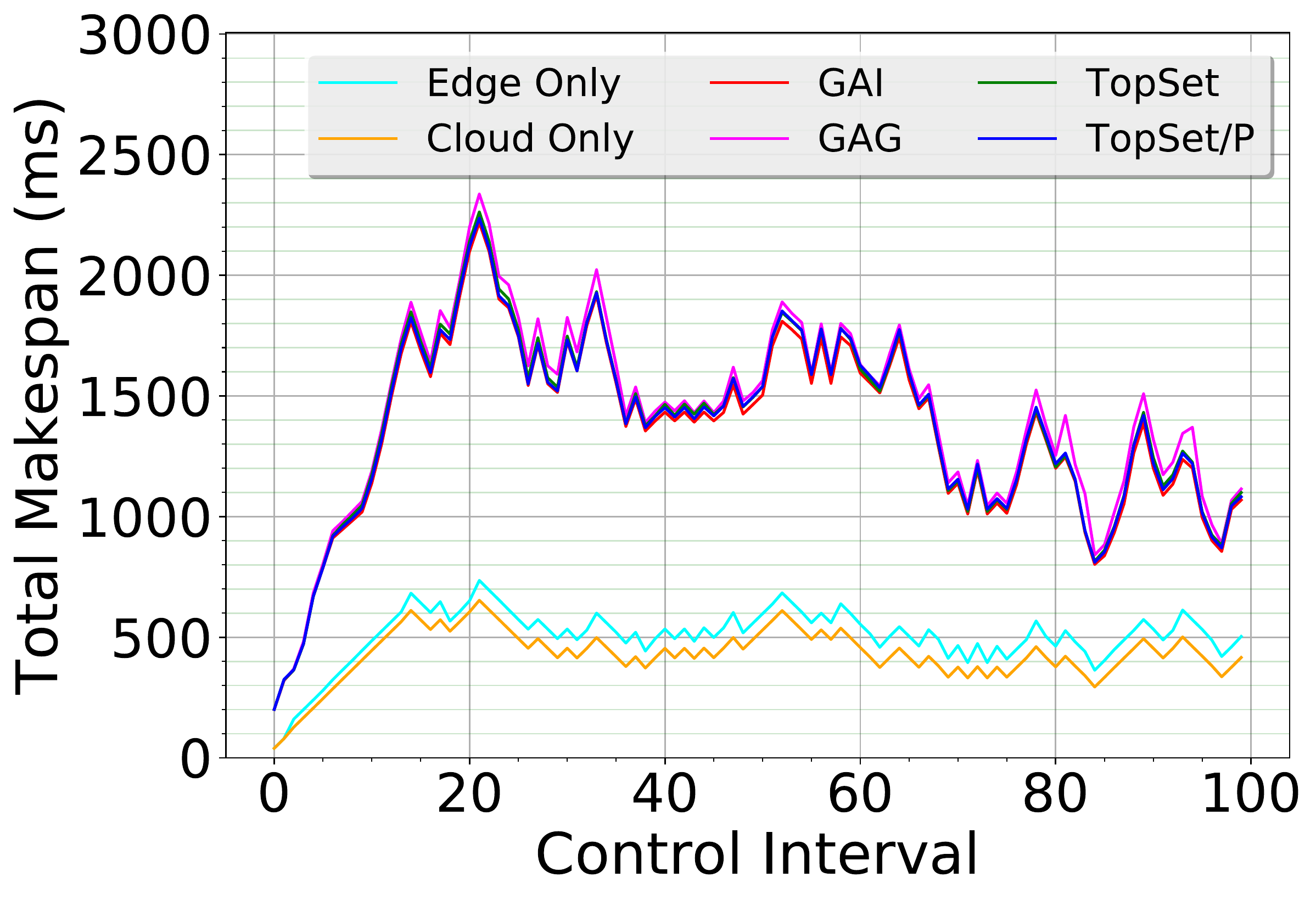}
		\label{fig:workload1:lat}
	}
	\subfloat{
		\includegraphics[width=0.23\textwidth]{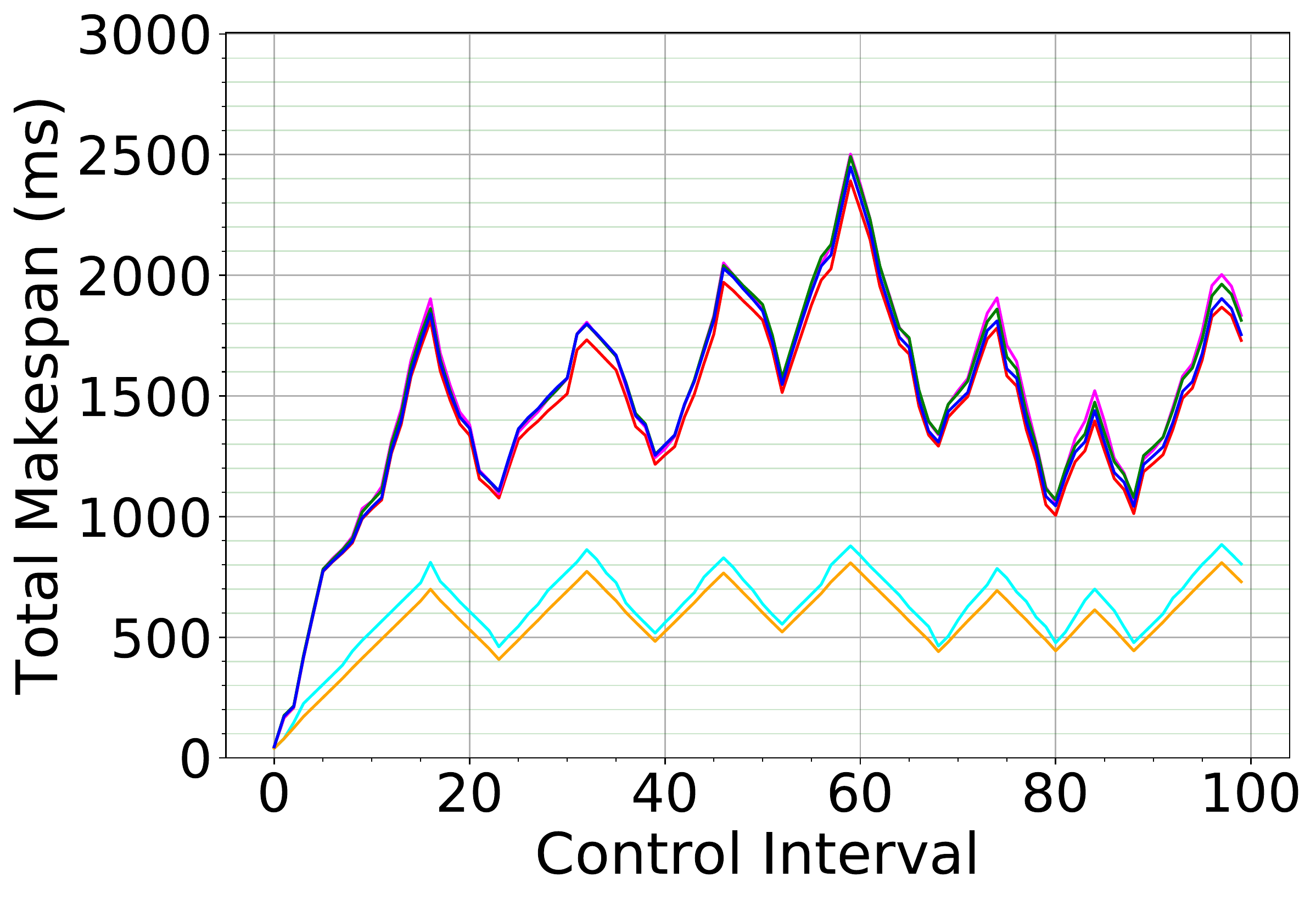}
		\label{fig:workload2:lat}
	}
	\subfloat{
		\includegraphics[width=0.23\textwidth]{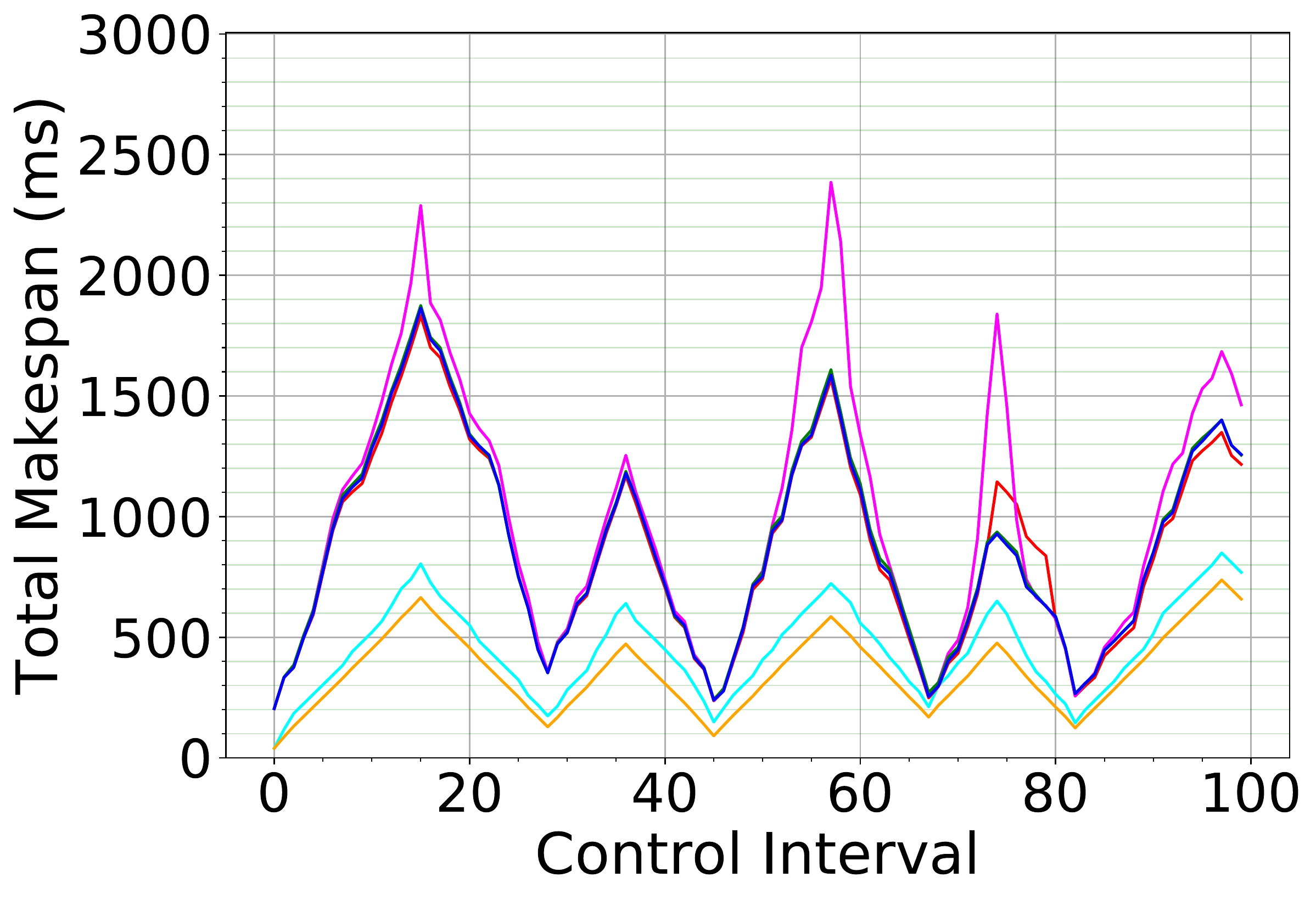}
		\label{fig:workload3:lat}
	}
	\subfloat{
		\includegraphics[width=0.23\textwidth]{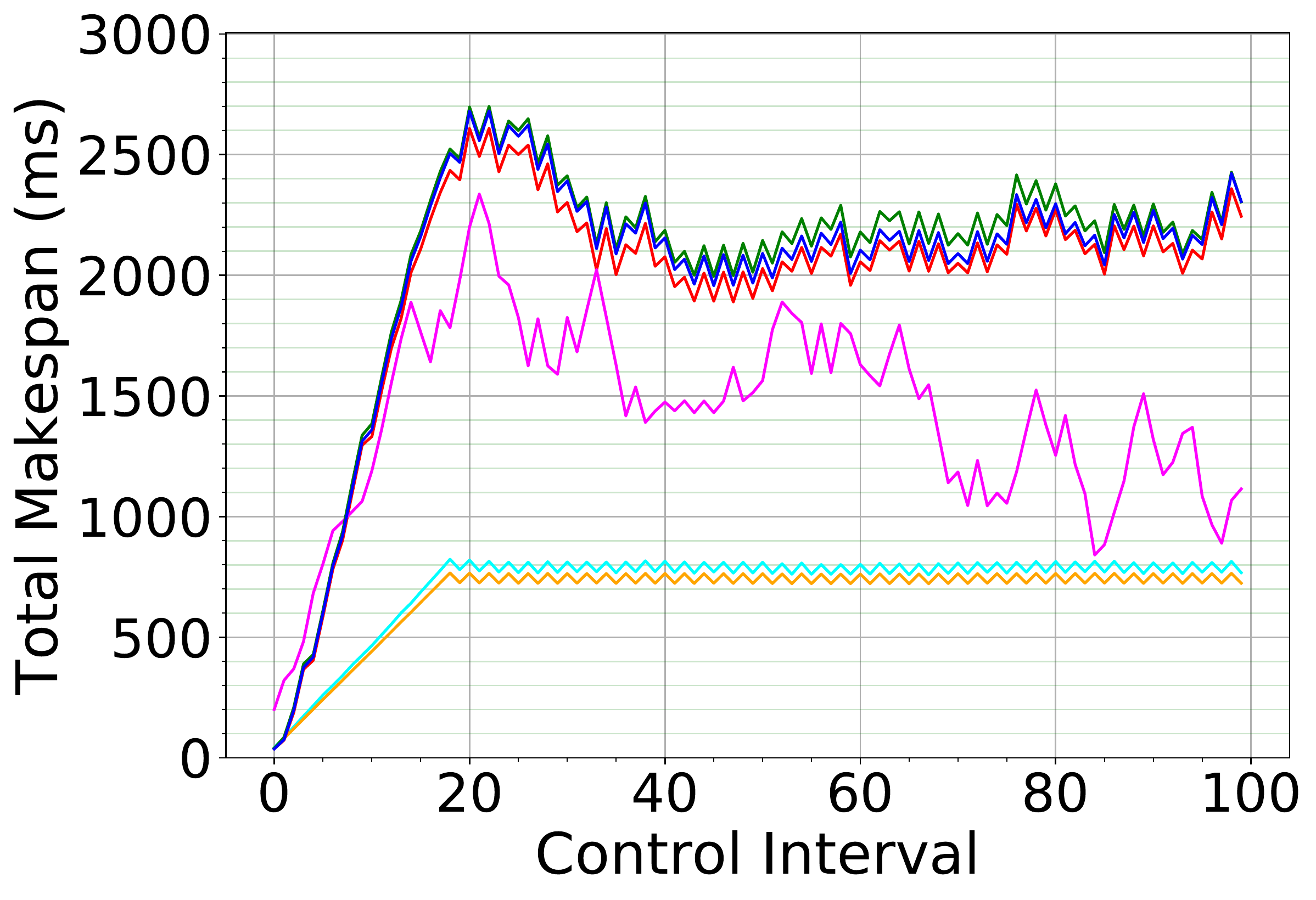}
		\label{fig:workload4:lat}
	}
       \subfloat{
		\footnotesize
		\begin{rotate}{270}\hspace{-1in}(i)~\emph{Small Setup}\end{rotate}
	}
\vspace{-0.1in}\\
\addtocounter{subfigure}{-5}
        \subfloat[Random Walk, $\widehat{U} = 2 \pm 0$]{
		\includegraphics[width=0.23\textwidth]{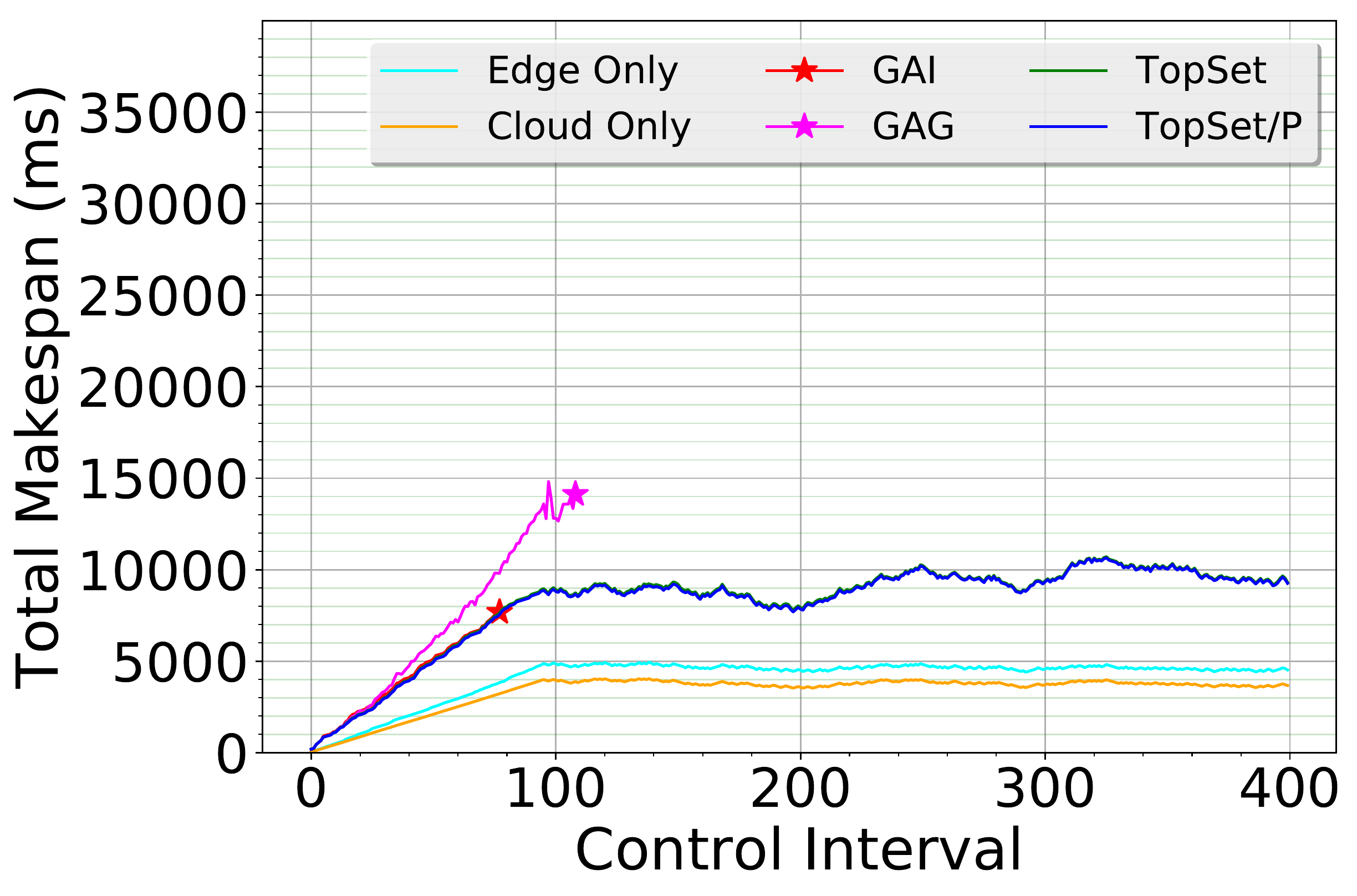}
		\label{fig:workload7:lat}
	}
	\subfloat[Random Walk, $\widehat{U} = 2 \pm 0.5$]{
		\includegraphics[width=0.23\textwidth]{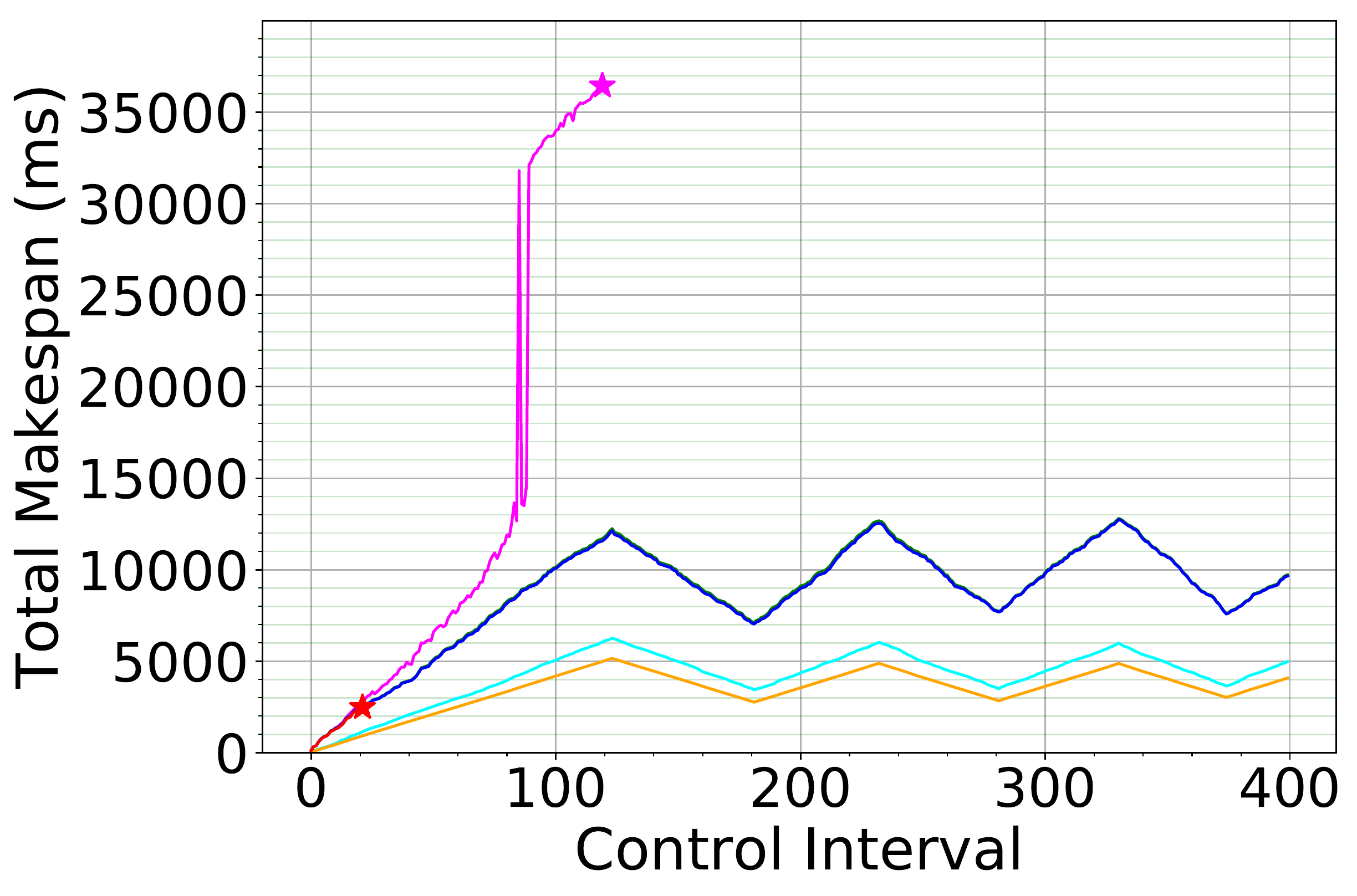}
		\label{fig:workload8:lat}
	}
	\subfloat[Random Walk, $\widehat{U} = 2 \pm 1.0$]{
		\includegraphics[width=0.23\textwidth]{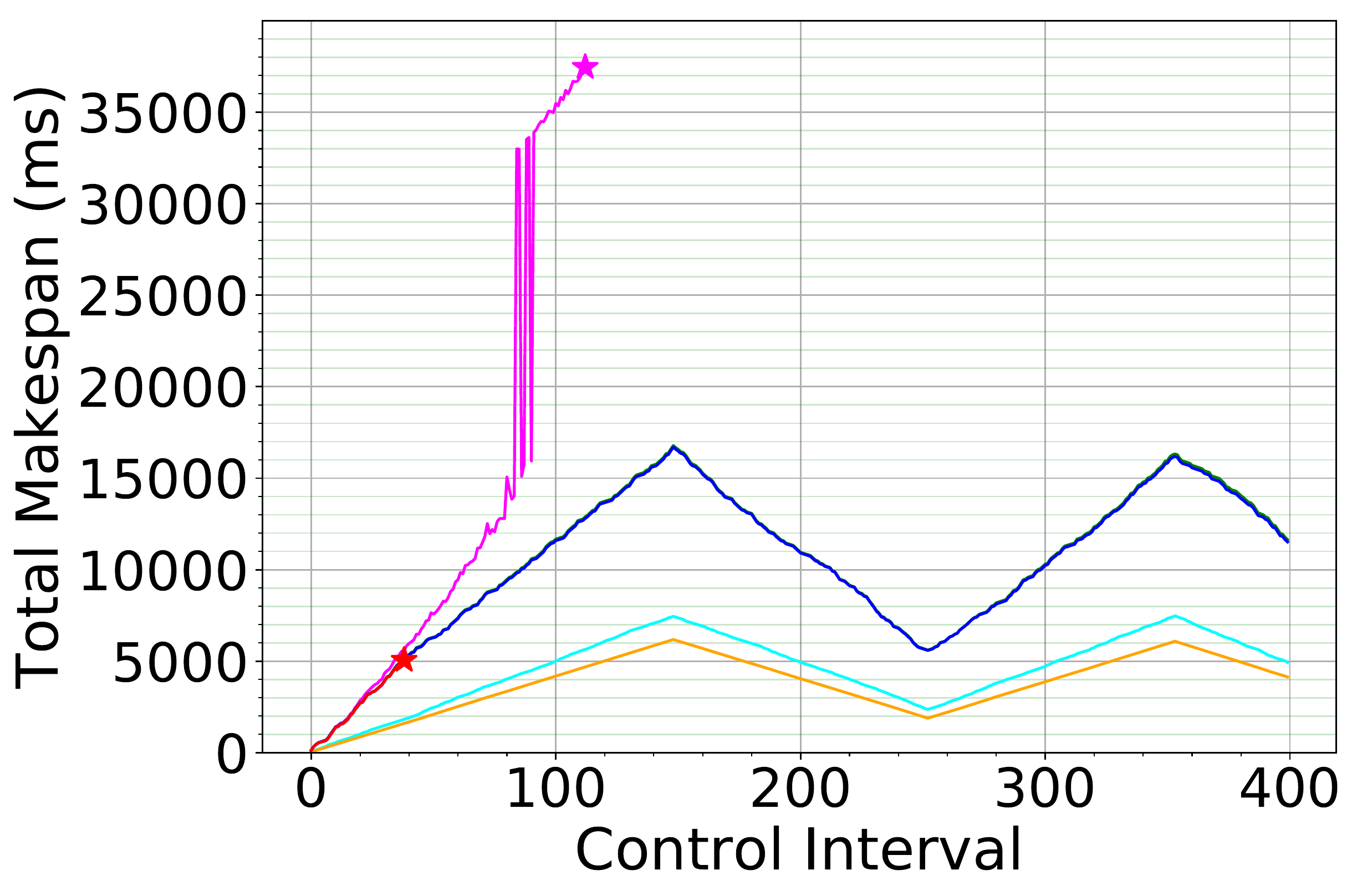}
		\label{fig:workload9:lat}
	}
	\subfloat[Poisson, $\lambda=12$]{
		\includegraphics[width=0.23\textwidth]{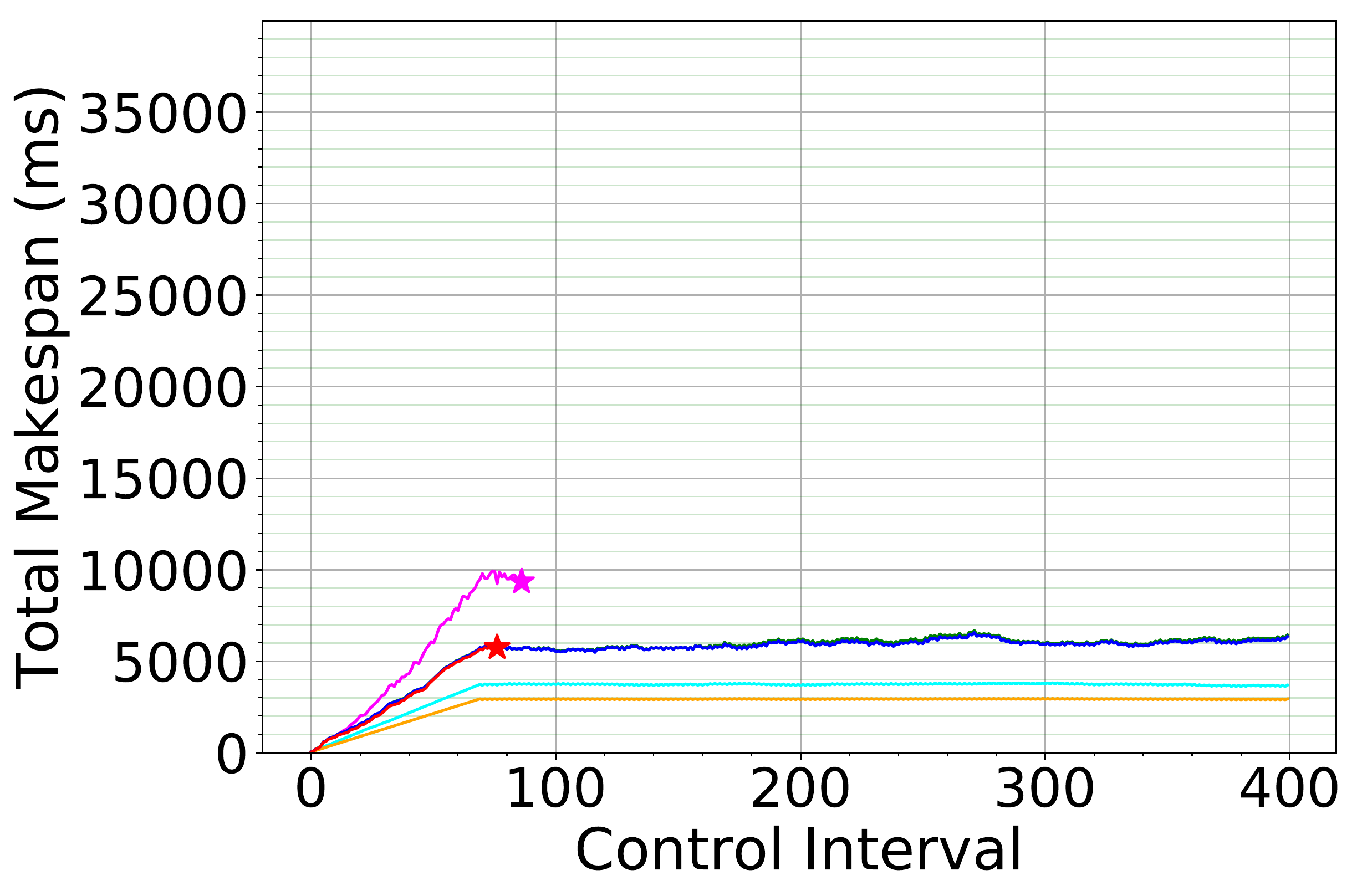}
		\label{fig:workload10:lat}
	}
	\subfloat{
		\footnotesize
		\begin{rotate}{270}\hspace{-1in}(ii)~\emph{Large Setup}\end{rotate}
	}
\vspace{-0.05in}
	\caption{\emph{Sum of DAG makespan} $\widehat{L}_{\mathbb{G}^{t+1}}$ at each control interval for placement strategies \emph{without rebalancing}}
	\label{fig:workload:lat}
\vspace{-0.15in}
\end{figure*}

\para{Quality of Overall Makespan}
Fig.~\ref{fig:workload:lat} show the sum of the end-to-end latency for all active DAGs (Y-Axis) for each control interval (X-Axis) for the $4$ workloads using the small (top row) and large (bottom row) setups, when \emph{not performing rebalance}. Besides our TopSet and TopSet/P heuristics, and the GAI and GAG meta-heuristics, we also test two na\"{i}ve baseline placement strategies of Edge-Only and Cloud-Only. 
In the \emph{Edge-Only baseline}, we consider an \emph{infinite availability} of edge resources with \emph{no network overhead} among them. Thus all non-sink queries are placed on \emph{exclusive} edges and the network latency from edge to Cloud VM is paid only once. In the \emph{Cloud-Only baseline}, we place all non-source queries on \emph{exclusive} Cloud VMs, again paying network latency only once from edge to Cloud. They do not consider constraint violations either. While \emph{infeasible}, these baselines offer a weak lower bound for makespan when not limited by resources and network latency.

In all the cases we see that the \textbf{\emph{Cloud-Only baseline}} has the least total makespan summed across all active DAGs since the compute speed of the Cloud resource is the fastest. \textbf{\emph{Edge-Only}} is the next best, and indicates the impact of the slower ARM CPU relative to the Xeon CPU for running queries. Neither take network time into account, and hence have a better makespan than all other heuristics. As expected, the large setup has a $10\times$ larger makespan ($\approx 5~secs$) than the small, due to $10\times$ more active vertices for the RW workloads. In case of Poisson workloads for large setup, it is 6x larger makespan. This is due to small DAGs having relatively higher probability of getting added unlike the RW workloads, thus contributing lesser makespan. 

For both the setups, \textbf{\emph{GAI}} performs better than \textbf{\emph{GAG}}, while we expect the global scheduling algorithm across all DAGs to better the incremental algorithm for just the new DAG. 
Upon examination, there are a large number of queries being scheduled by GAG in both the setup, especially as the utilization peaks. This causes the search space to explode, resulting in a sub-optimal solution. For the large setup, this space explosion is seen for both GAI and GAG, and the strategies fail to return any solution beyond $\approx 100$ intervals (`$\star$' in Fig.~\ref{fig:workload:lat}(ii)). GAG fails to return a solution at an earlier control interval than GAI. Tuning the GA meta-parameters did not help either. 
\ysnoted{Why is GAG much better for poisson? Any errors?} 


We see that our \textbf{\emph{TopSet and TopSet/P}} heuristics have total makespan values that match closely with the better of GAI or GAG, and they are able to find a solution in \emph{all} cases. This exhibits the robustness of our heuristics. 
We also notice that TopSet/P is marginally but consistently better than TopSet. The latter only considers the current query being visited in the topological set traversal to find the cost of mapping to a resource, while the former also estimates the side-effect of the mapping on other queries with a penalty function. We see the benefit of that optimization here. 
\ysnoted{\ysnote{Are all GA algorithms giving a valid solution? Do the peaks in W/L 3 for GAA and step 75 of GAI indicate that they violate constraints? Verify timeline plots for rebalance}\drnote{Rebalance plots also show similar step 75 for W/L3 GAI}}
%

\begin{figure*}[t!]
	\centering
\subfloat{
		\includegraphics[width=0.23\textwidth]{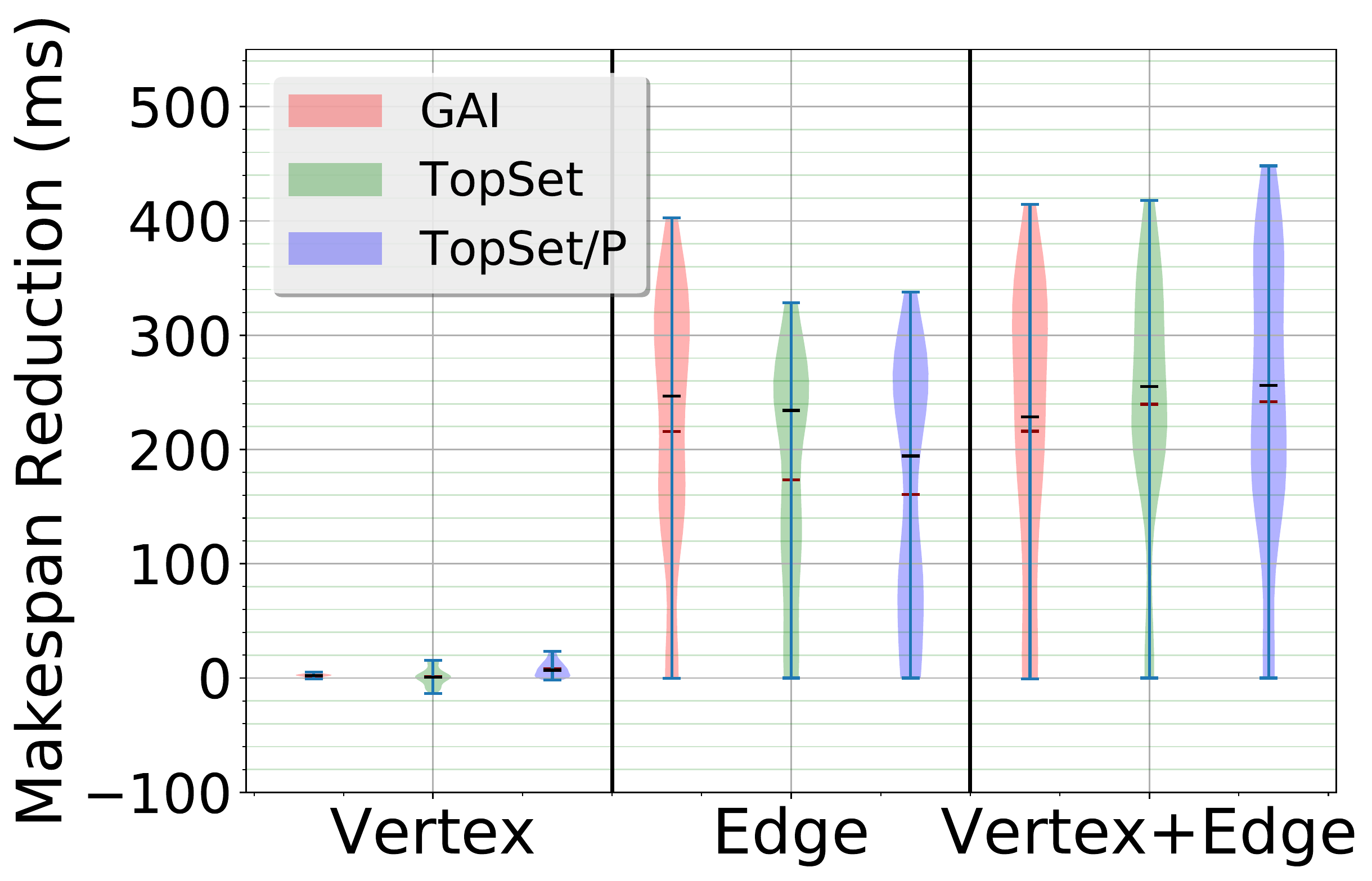}
		\label{fig:workload1:rebalance}
	}
	\subfloat{
		\includegraphics[width=0.23\textwidth]{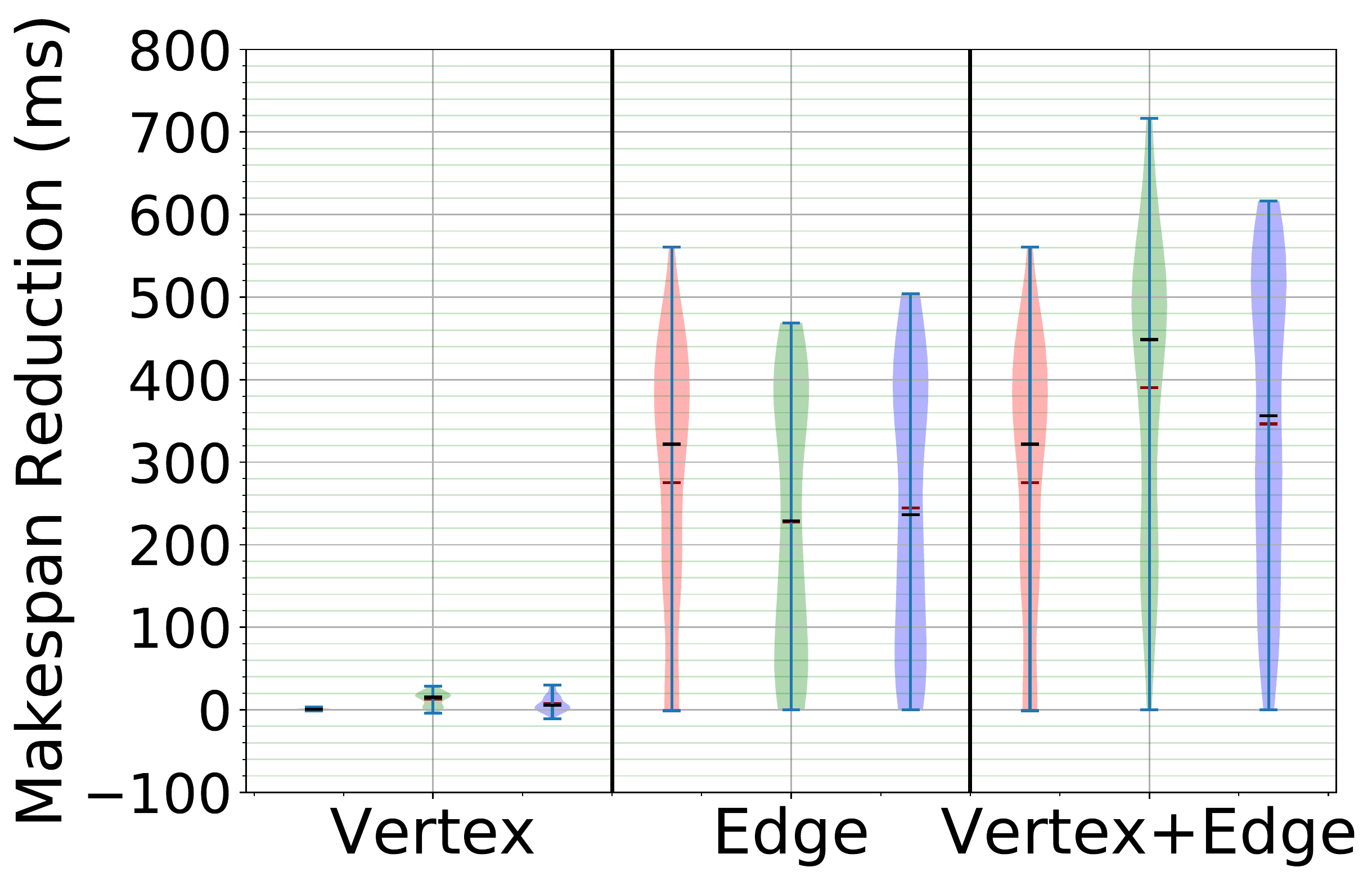}
		\label{fig:workload2:rebalance}
	}
	\subfloat{
		\includegraphics[width=0.23\textwidth]{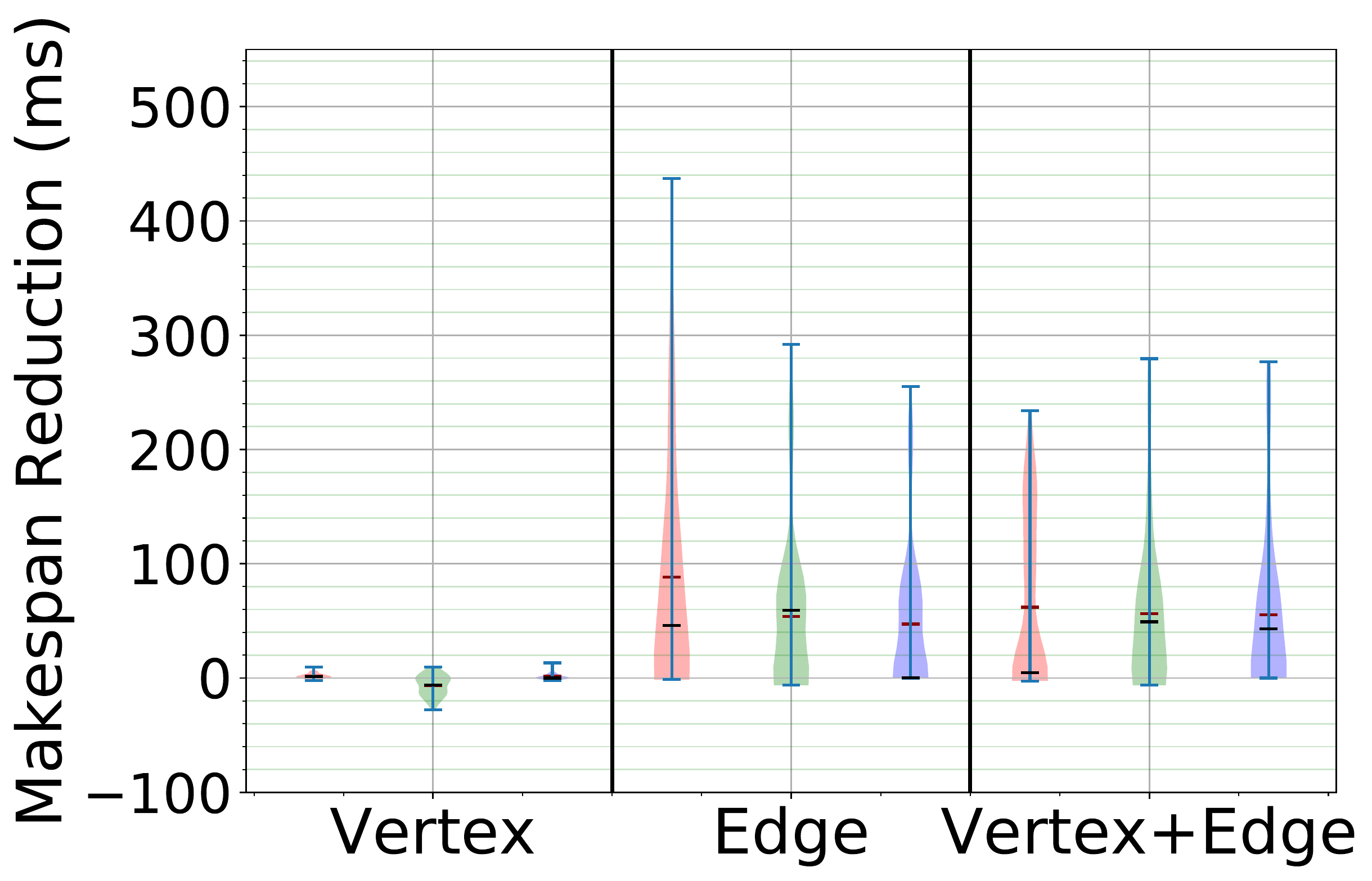}
		\label{fig:workload3:rebalance}
	}
	\subfloat{
		\includegraphics[width=0.23\textwidth]{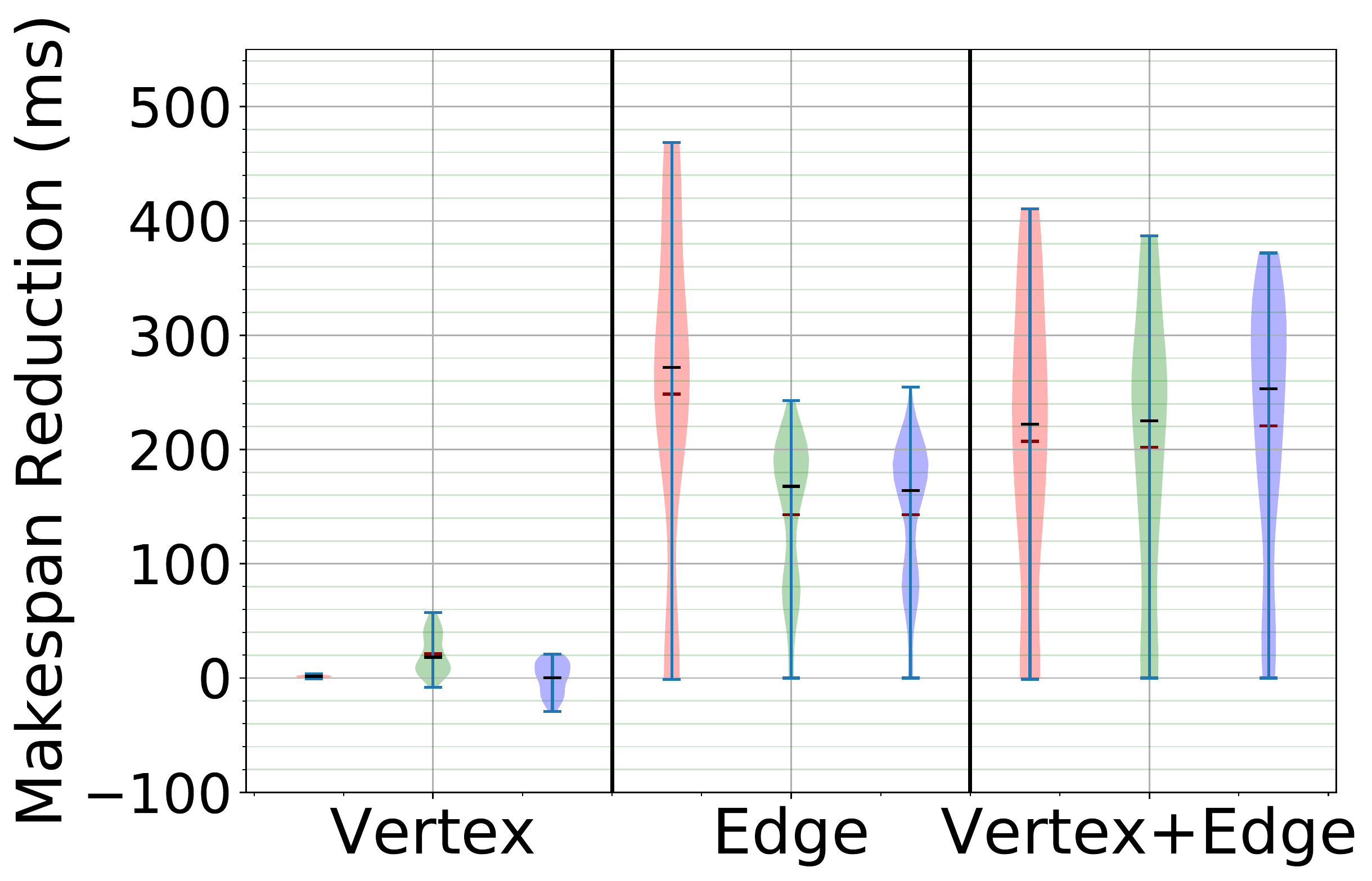}
		\label{fig:workload4:rebalance}
	}
        \subfloat{
		\footnotesize
		\begin{rotate}{270}\hspace{-1in}(i)~\emph{Small Setup}\end{rotate}
	}
\vspace{-0.1in}
\\
\addtocounter{subfigure}{-5}
\subfloat[Random Walk, $\widehat{U} = 2 \pm 0$]{
	\includegraphics[width=0.23\textwidth]{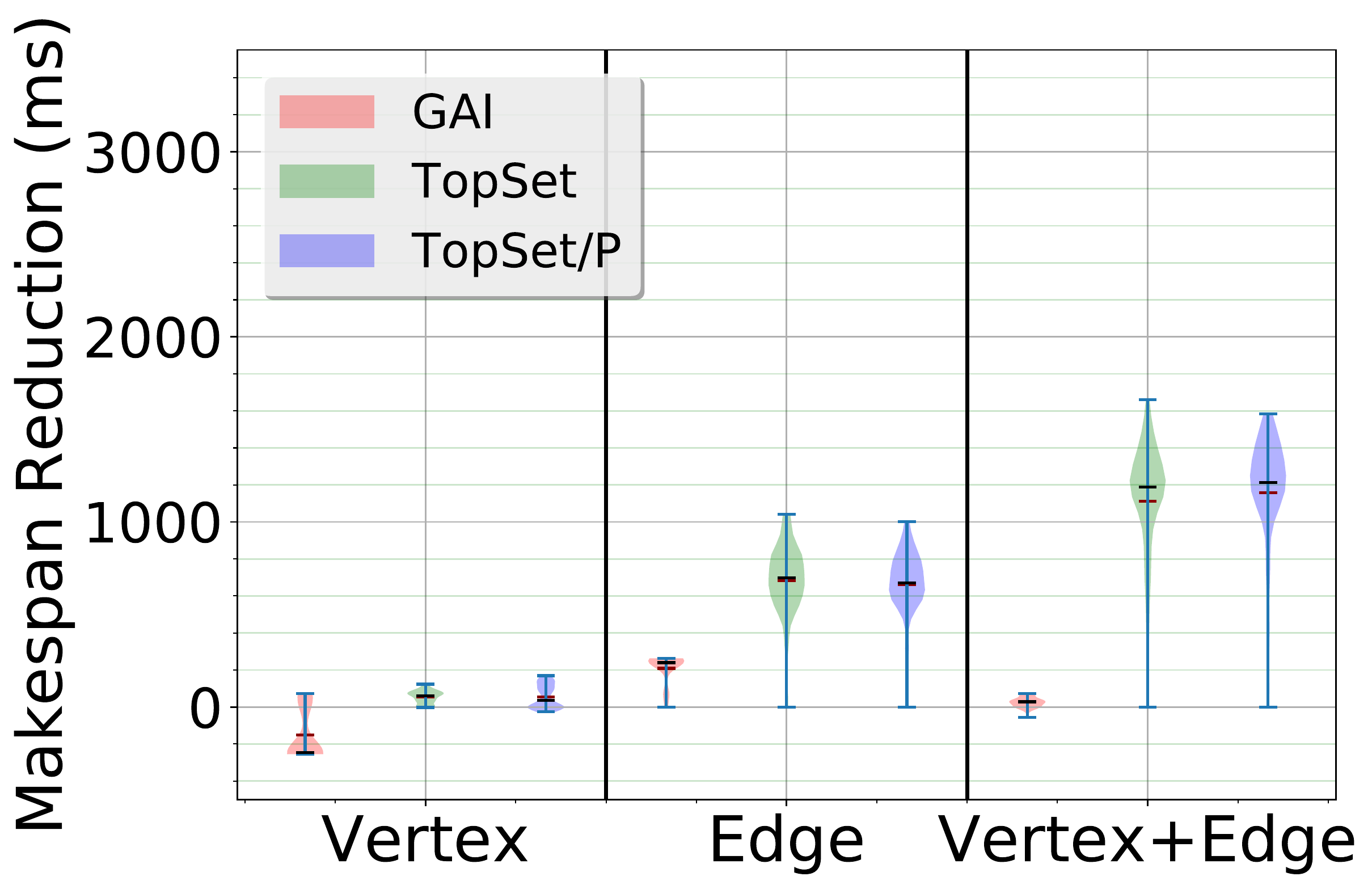}
	\label{fig:workload7:rebalance}
}
\subfloat[Random Walk, $\widehat{U} = 2 \pm 0.5$]{
	\includegraphics[width=0.23\textwidth]{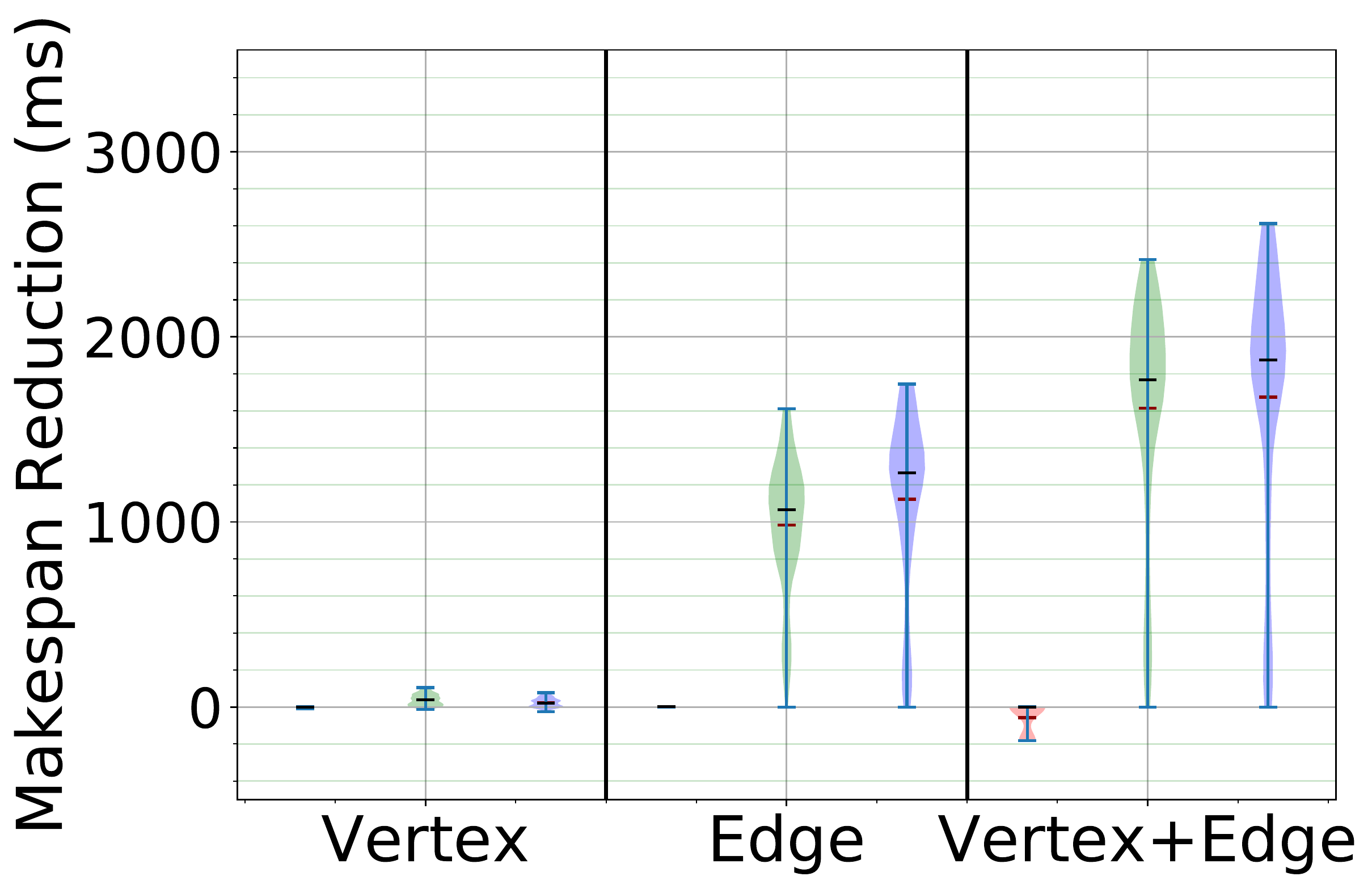}
	\label{fig:workload8:rebalance}
}
\subfloat[Random Walk, $\widehat{U} = 2 \pm 1.0$]{
	\includegraphics[width=0.23\textwidth]{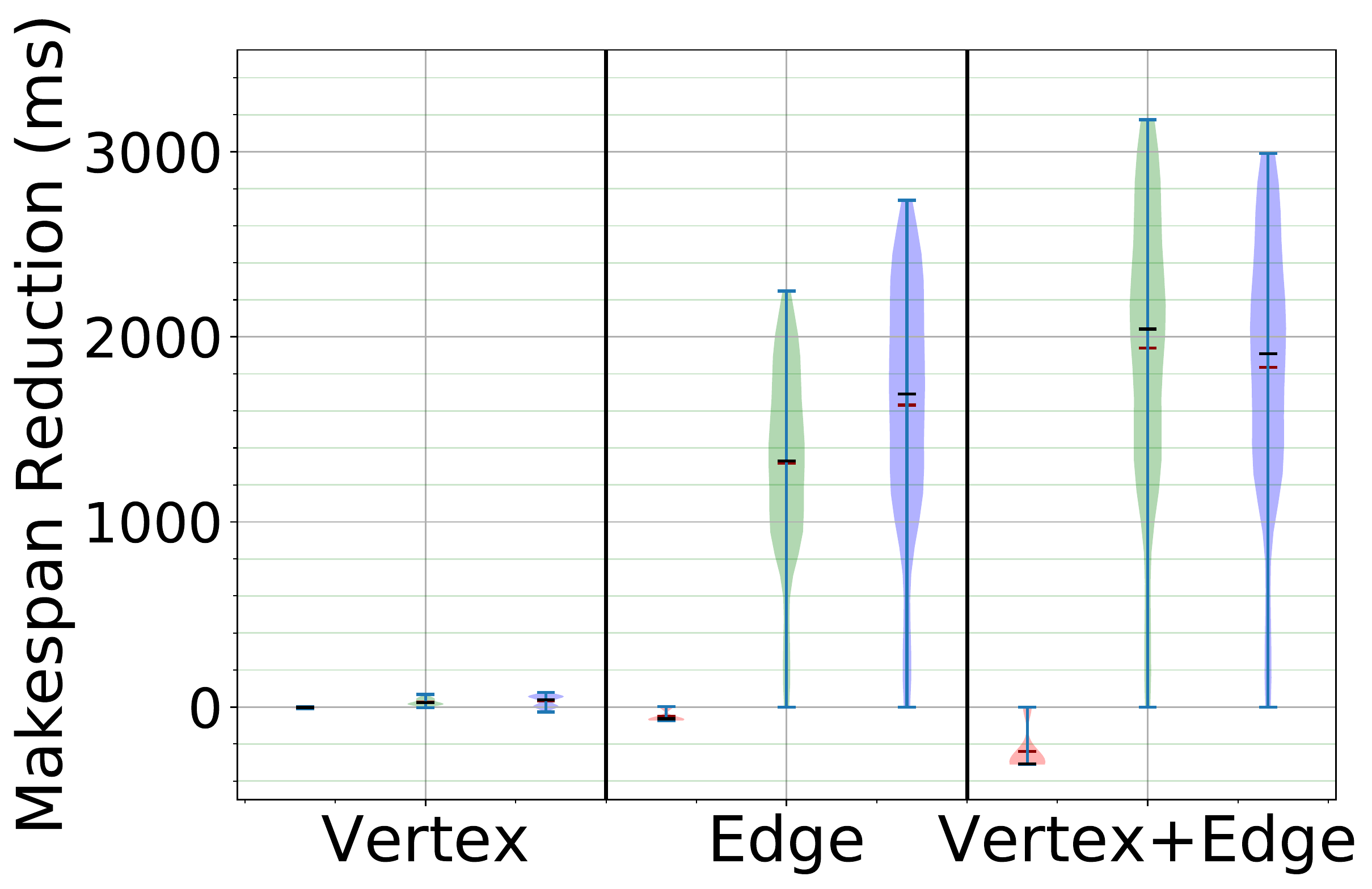}
	\label{fig:workload9:rebalance}
}
\subfloat[Poisson, $\lambda=12$]{
	\includegraphics[width=0.23\textwidth]{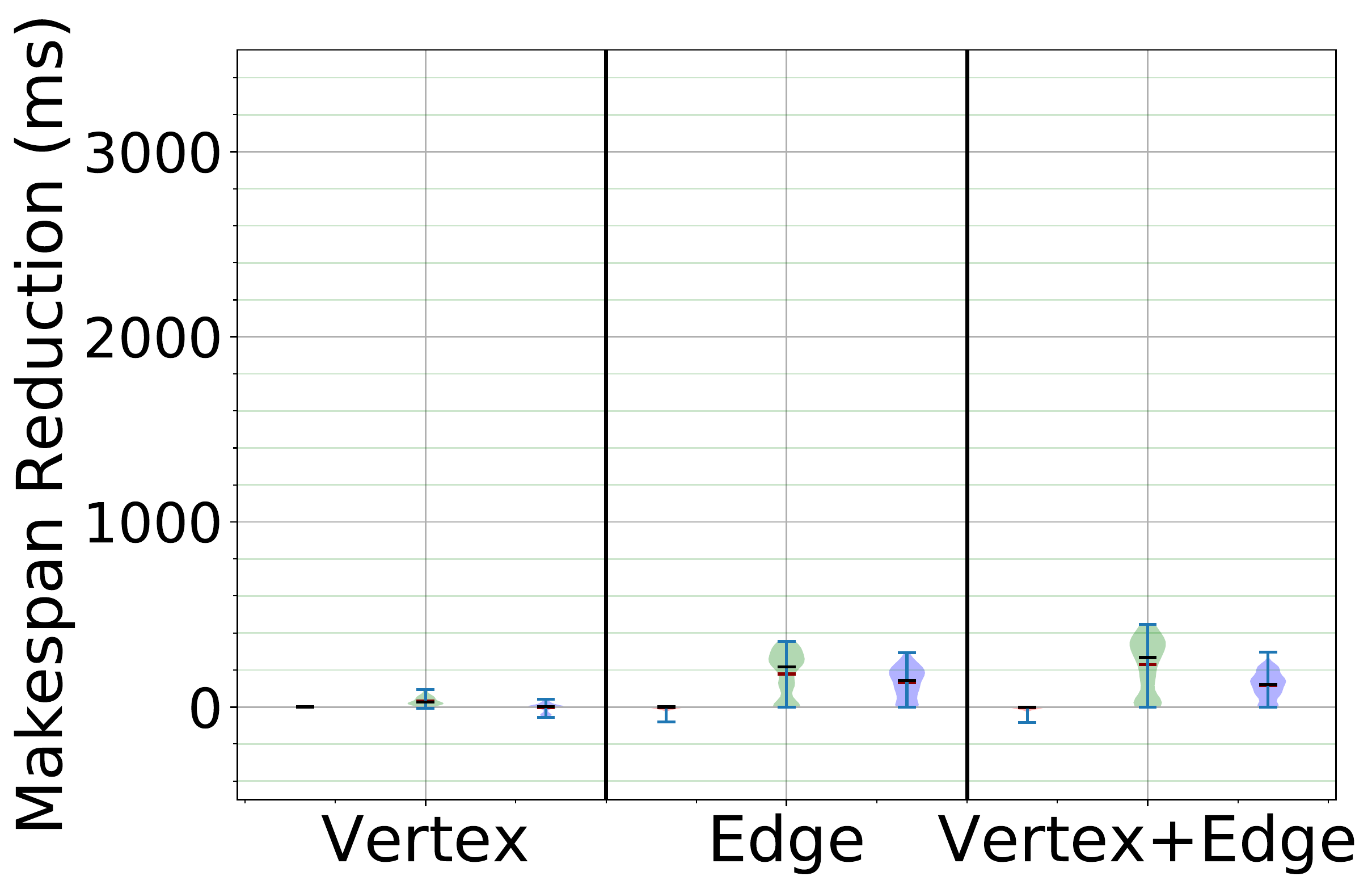}
	\label{fig:workload10:rebalance}
}
\subfloat{
	\footnotesize
	\begin{rotate}{270}\hspace{-1in}(ii)~\emph{Large Setup}\end{rotate}
}
\vspace{-0.05in}
\caption{Violin plot of \emph{reduction in sum of makespan} with different \emph{rebalance strategies}, relative to no rebalance (Fig.~\ref{fig:workload:lat}). }
\label{fig:workload:rebalance}
\vspace{-0.15in}
\end{figure*}

We measure the impact of the \textbf{\emph{rebalance}} strategies that complement the heuristics after their initial placement. The Fig.~\ref{fig:workload:rebalance} plot the relative improvement in makespan when performing rebalance after arrival and departure of DAGs, compared to without. A positive value indicates a reduction in total makespan. Rebalance is only relevant to TopSet, TopSet/P and GAI, since GAG does a global DAG placement. GAI values are less relevant for the large setup due to few valid solutions. 
While \emph{vertex rebalance} appears to have only a small impact, sometimes even doing worse, the \emph{edge rebalance} has significant benefits, reducing the total makespan by up to $20\%$ for the small setup and $25\%$ for the large setup. This confirms that network plays a greater role in the DAG latency in such distributed setups. Interestingly, for the Poisson and RW $2 \pm 0.5$ workloads, combining \emph{vertex and edge rebalance} is better than just edge even though vertex rebalance by itself offers limited benefits. While all three heuristics see benefits from rebalance, GAI for the small setup appears to improve more frequently as it starts from a better solution base. 



\begin{figure}[t]
	\centering
	\subfloat[Small Setup]{
		\includegraphics[width=.45\columnwidth]{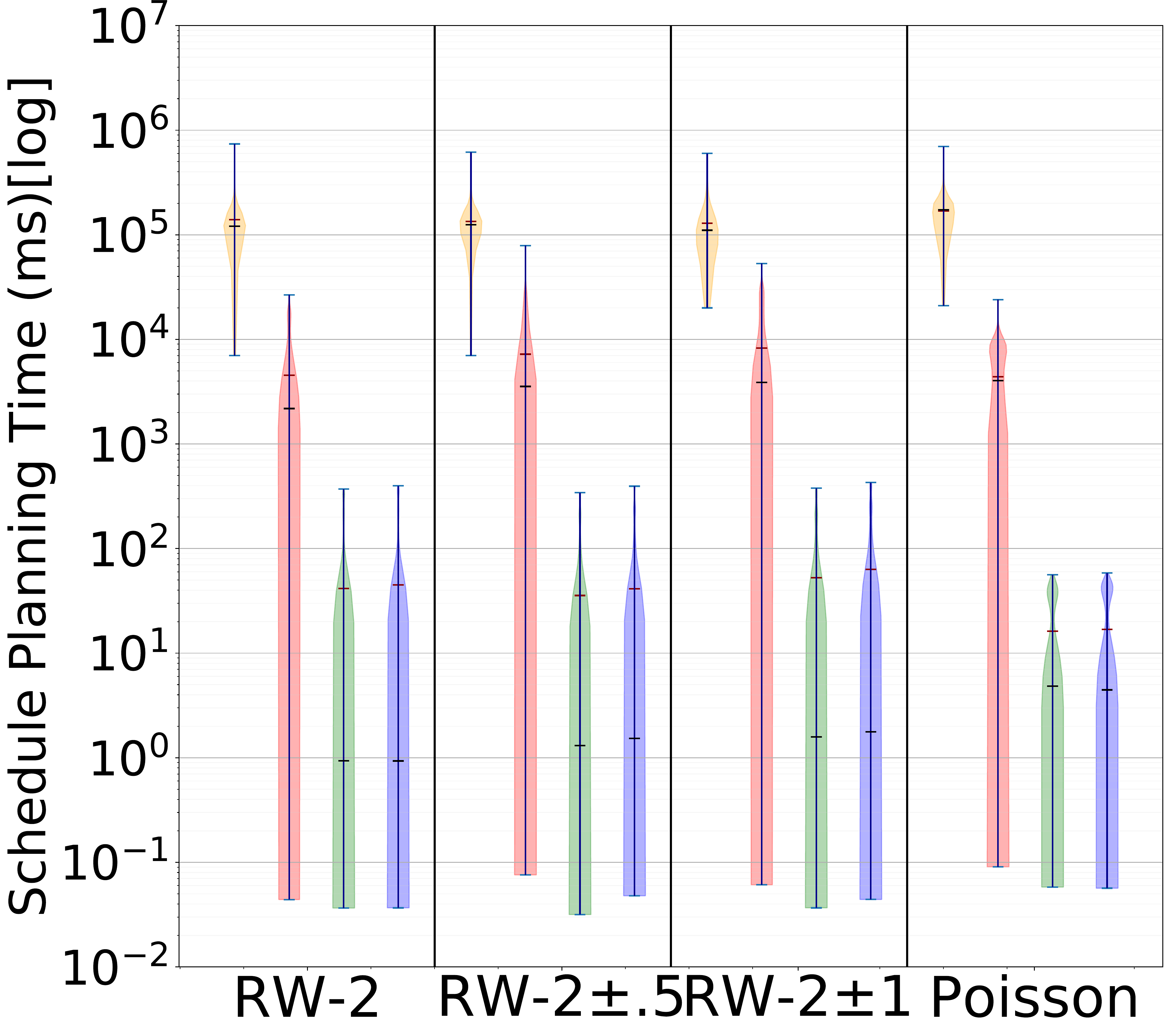}
	}
		\subfloat[Large Setup]{
			\includegraphics[width=.45\columnwidth]{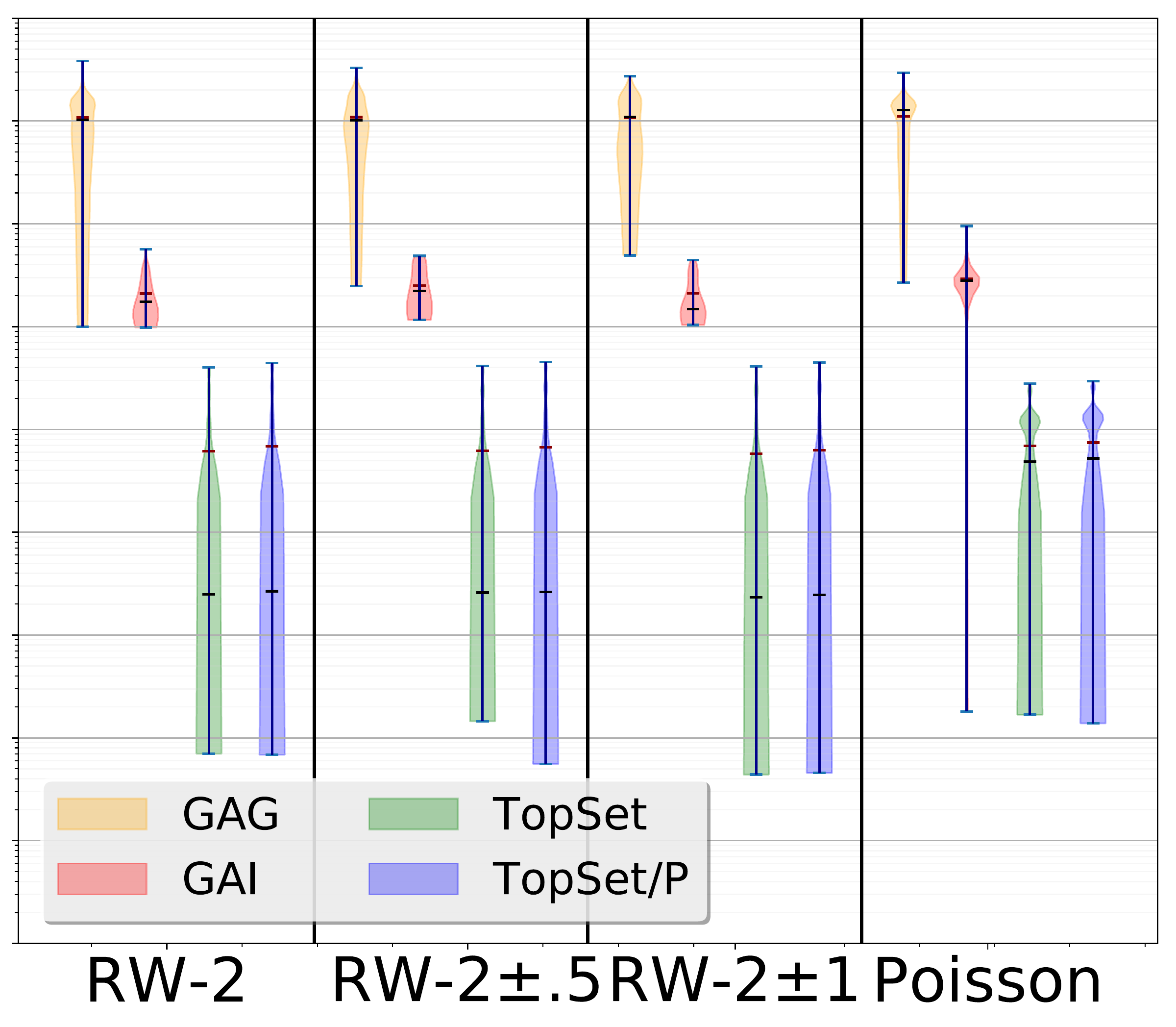}
		}
\vspace{-0.05in}
	\caption{Violin plot of \emph{Schedule Planning time}, $\phi_t$ for each workload and strategy}
	\label{fig:workload:scheduling}
\vspace{-0.15in}
\end{figure}
\para{Runtime Performance} The secondary measures of the solution quality have to do with reducing the schedule planning time, number of migrations and stabilization time.
Fig.~\ref{fig:workload:scheduling} shows the \textbf{\emph{schedule planning time}} taken by the four heuristic algorithms upon DAG arrival (and removal too for GAG) at a control interval. The Y-axis is in log scale. We see that the GA-based meta-heuristics take much more time to converge to a placement than our TopSet strategies. In fact, our heuristics give a placement mapping within $1~sec$ in all cases for all workloads, barring some outliers. GAG and GAI take $10^4$ and $10^3$ longer, ranging from under a minute for GAI to multiple hours for GAG. 
We also see (not shown) that there is strong correlation between the number of queries being scheduled and the time taken for the schedule planning, which is also correlated with utilization. As a result, TopSet and TopSet/P are not just robust in always giving a valid solution and matching the GA's quality when they return a solution, but also do so in sub-second time. This makes them well-suited for short control intervals where one can expect $1-10$ DAGs change every second. The GA solutions are not feasible for large setups, and only GAI is practically usable for small setups. Failure of such meta-heuristics to scale with parameter space in terms of solution quality and run-time performance is evident in~\cite{Silva:2016}. 



\begin{figure*}[t]
	\centering
	\subfloat[Random Walk, $\widehat{U} = 2 \pm 0$]{
		\includegraphics[width=0.23\textwidth]{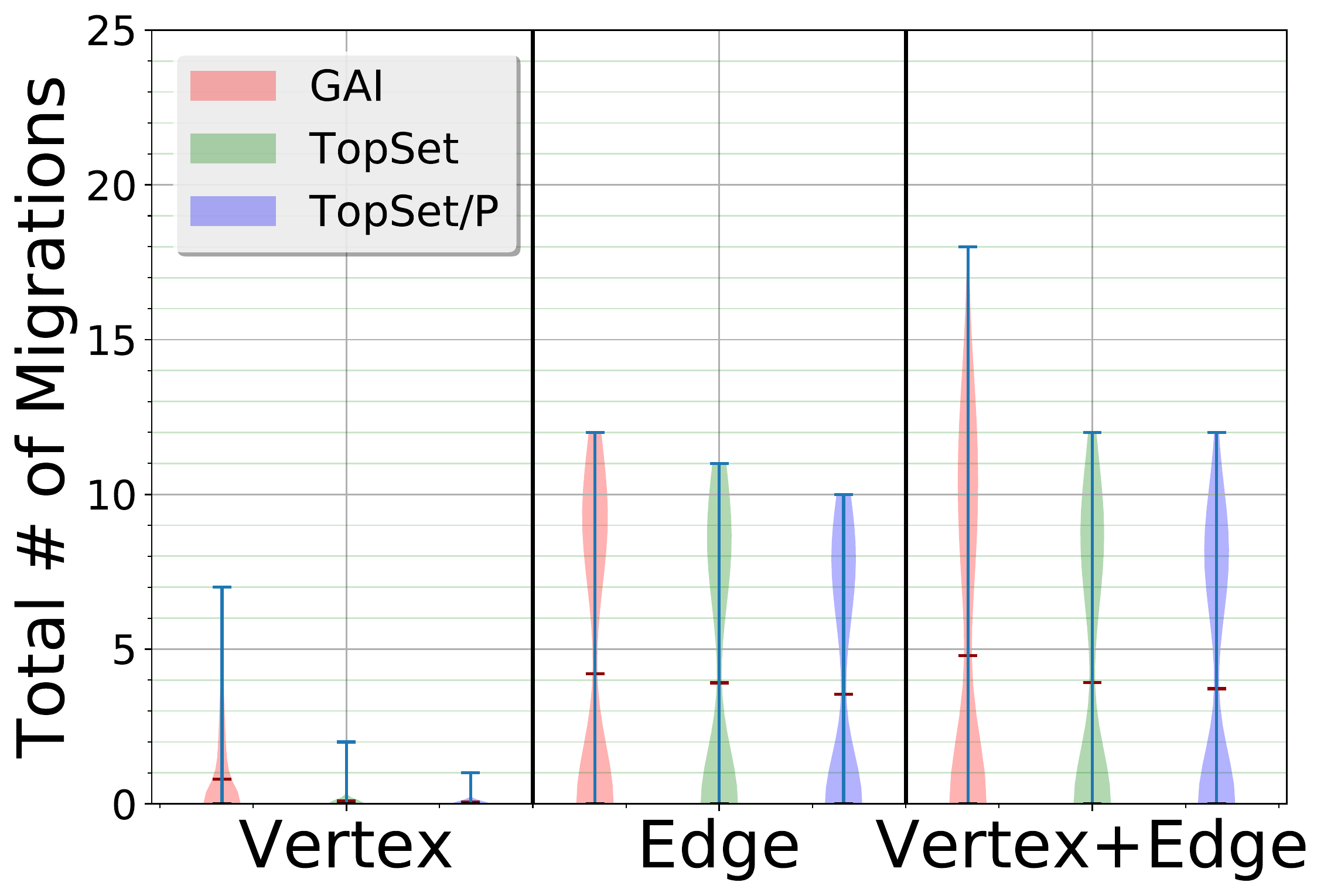}
		\label{fig:workload1:mig_Violin}
	}
	\subfloat[Random Walk, $\widehat{U} = 2 \pm 0.5$]{
		\includegraphics[width=0.23\textwidth]{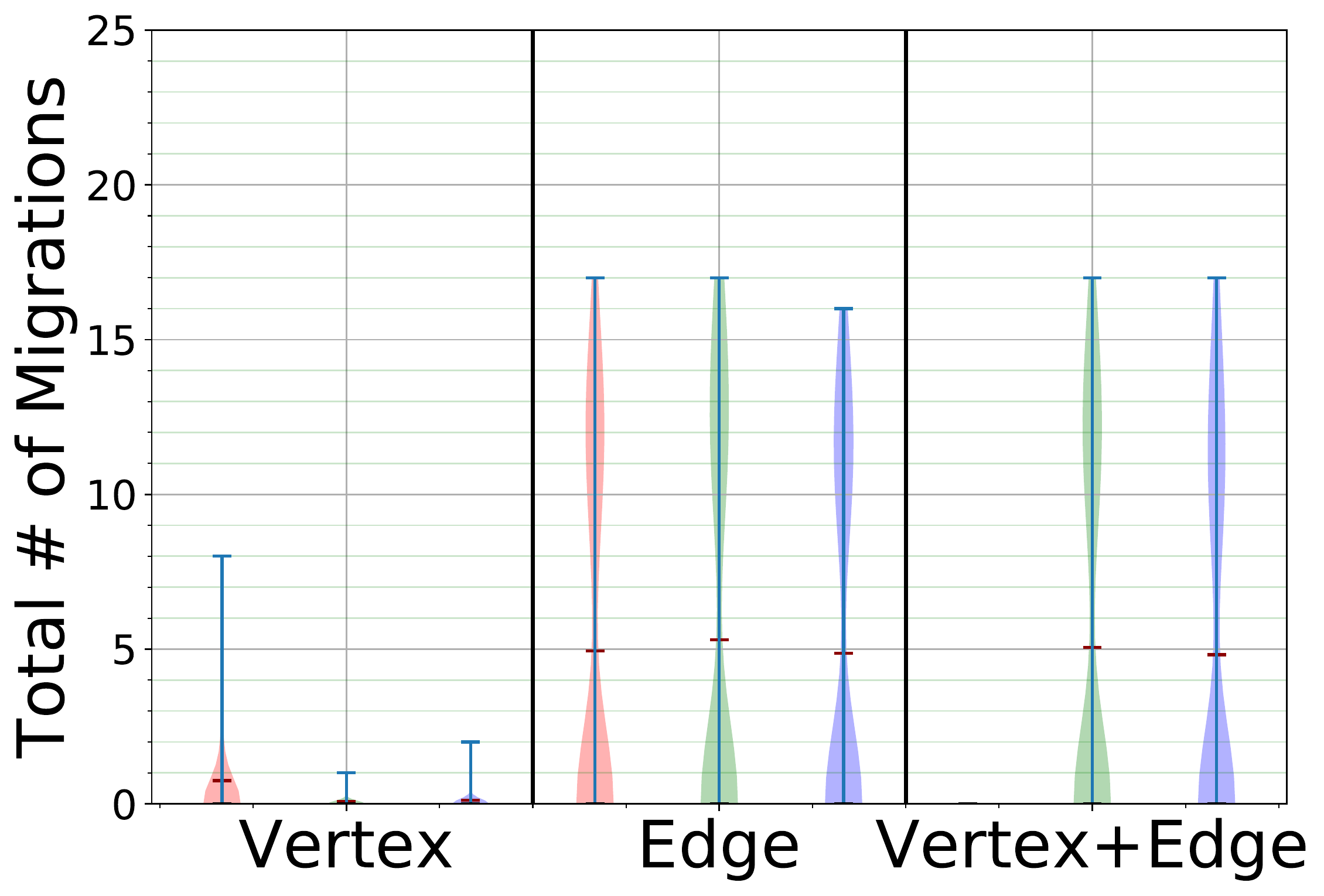}
		\label{fig:workload2:mig_Violin}
	}
	\subfloat[Random Walk, $\widehat{U} = 2 \pm 1.0$]{
		\includegraphics[width=0.23\textwidth]{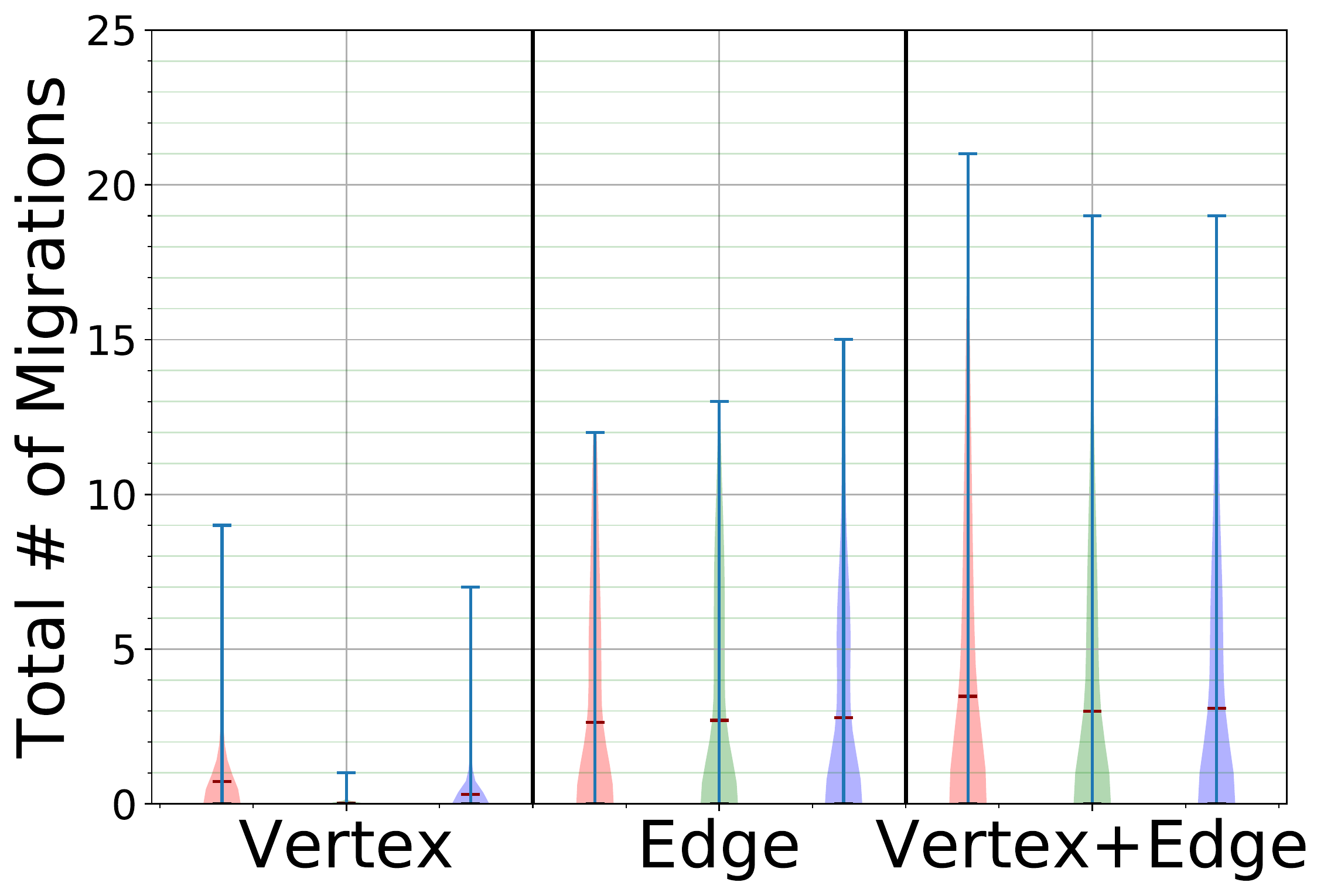}
		\label{fig:workload3:mig_Violin}
	}
		\subfloat[Poisson, $\lambda=12$]{
			\includegraphics[width=0.23\textwidth]{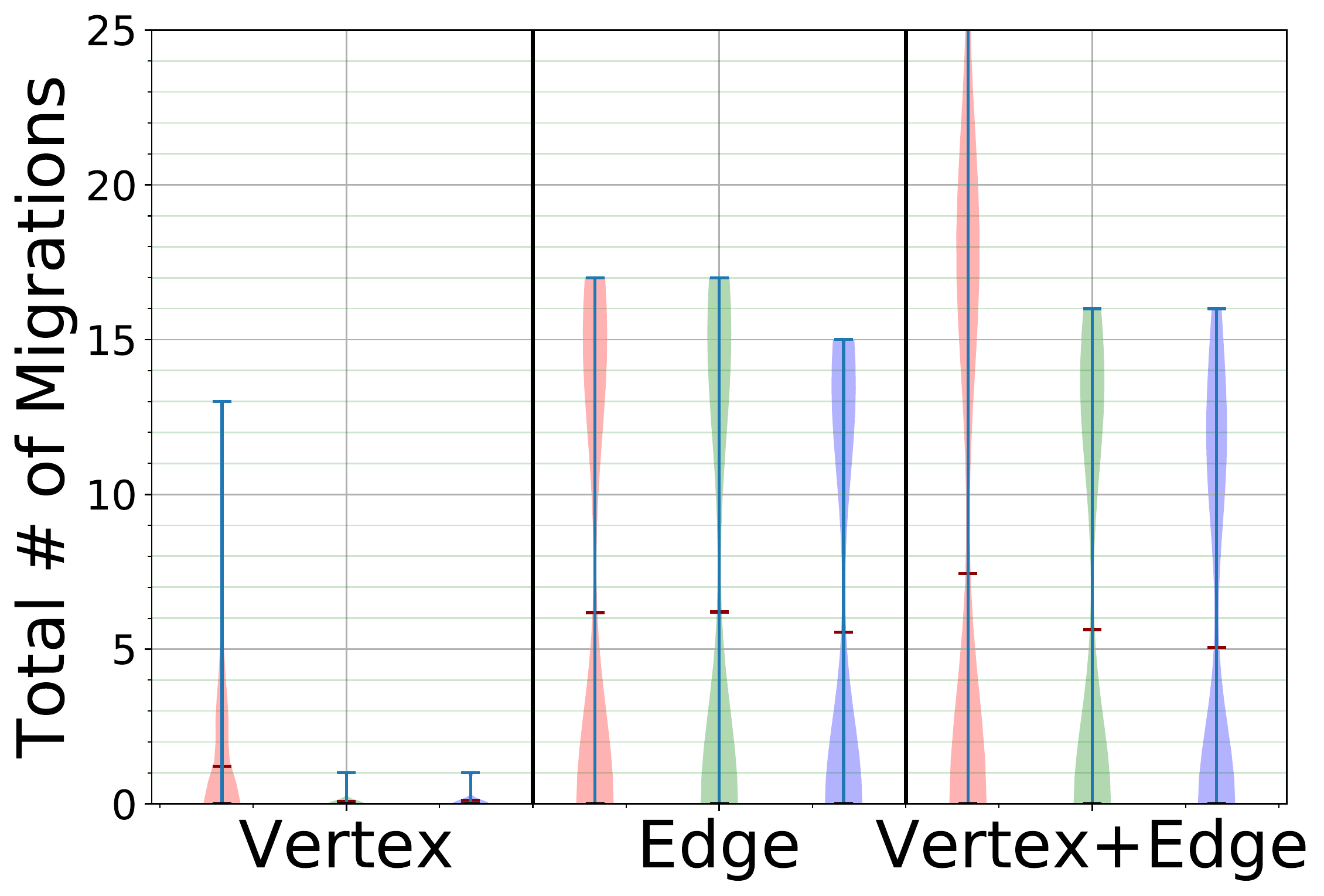}
			\label{fig:workload4:mig_Violin}
		}
\vspace{-0.05in}
	\caption{\emph{Number of migrations} $\overline{\rho_t}$ required for different placement strategies with rebalancing, for Small setup}
	\label{fig:workload:migration_Violin}
\vspace{-0.15in}
\end{figure*}
We also report the \textbf{\emph{number of migrations}} performed by the strategies for different workloads of the small setup in Figs.~\ref{fig:workload:migration_Violin}. We omit the plots for large setup due to space limits. We see that vertex rebalancing causes fewer migrations compared to edge rebalancing. While good, this has the consequence of having minimal impact on improving the makespan after applying rebalancing. Edge rebalancing on the other hand has more migrations but also a better makespan reduction. 
Further, the median number of migrations in all cases is zero, indicating that migrations are less frequent and rebalance is able to cause an improvement in $<50\%$ of the schedules. For the large setup, we see up to $140$ migrations take place compared to a peak of about $25$ for the small setup. More resources imply a larger solution space for improvements. While not shown, GAG causes most queries in the global set of DAGs to migrate each time. 

\begin{figure}[t!]
	\centering
	\subfloat[RW, $2 \pm 0$]{%
		\includegraphics[width=0.23\columnwidth]{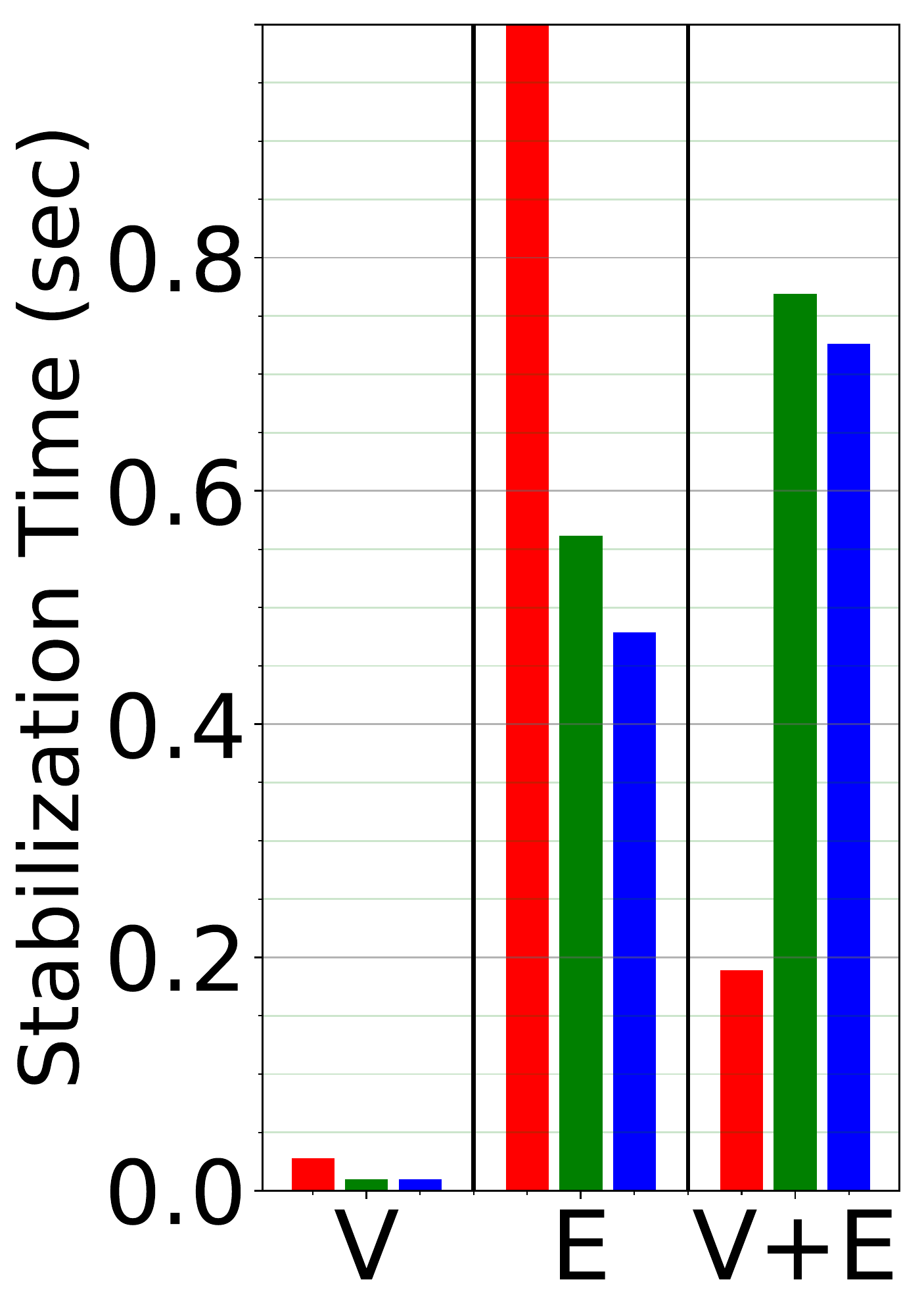}%
		\label{fig:workload1:stab_Violin}
	}
	\subfloat[RW, $2 \pm 0.5$]{%
		\includegraphics[width=0.227\columnwidth]{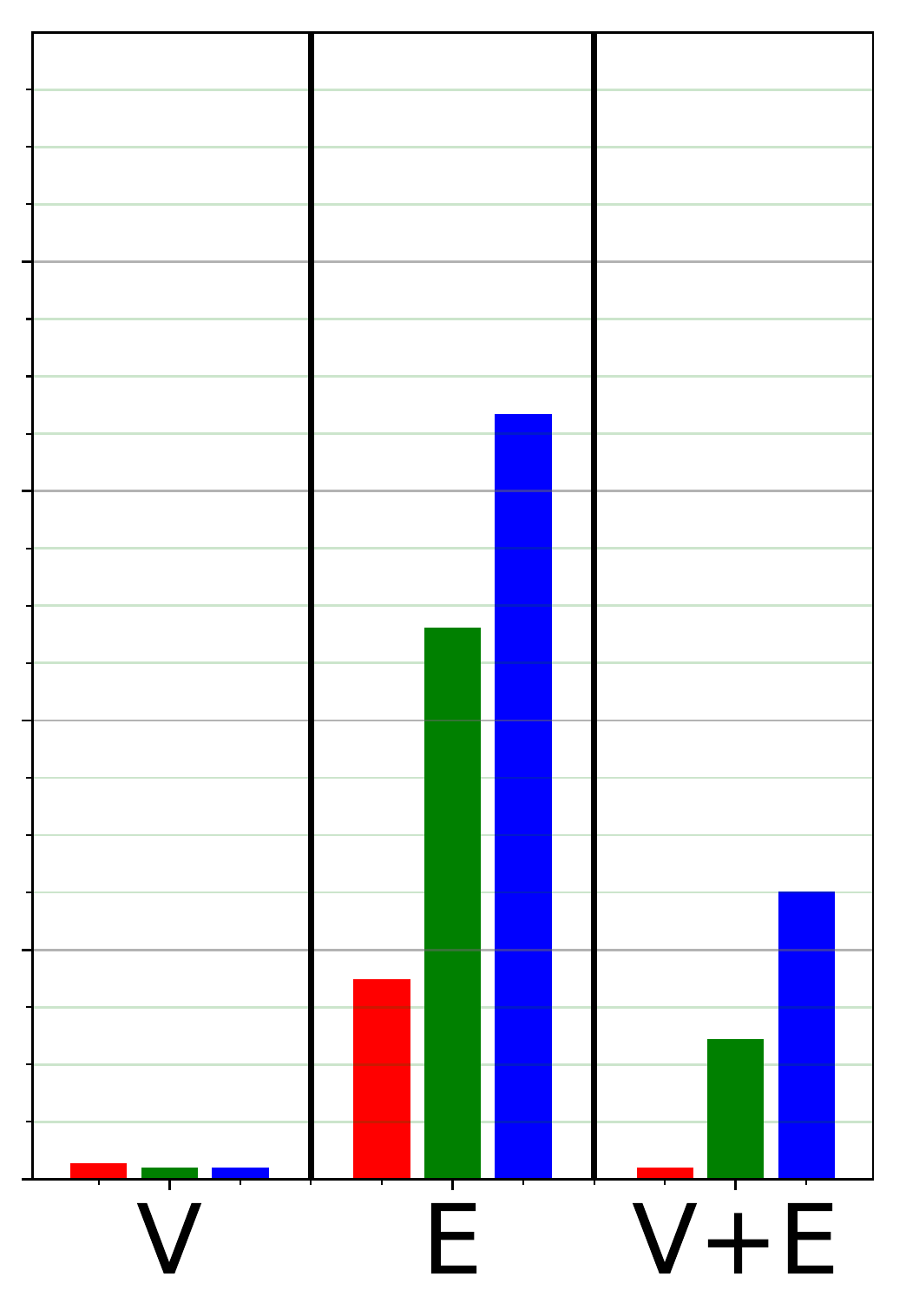}%
		\label{fig:workload2:stab_Violin}
	}
	\subfloat[RW, $2 \pm 1.0$]{%
		\includegraphics[width=0.227\columnwidth]{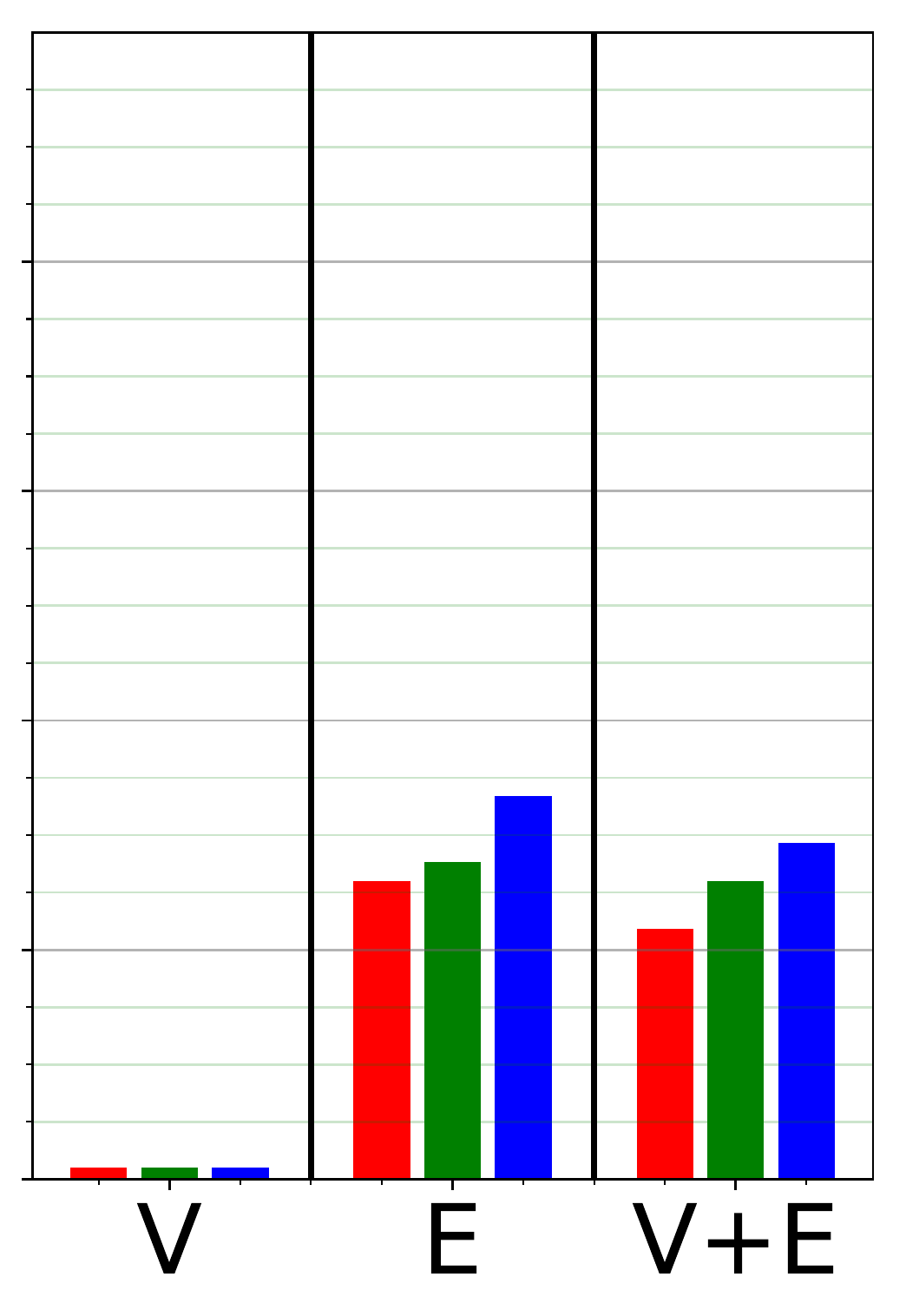}%
		\label{fig:workload3:stab_Violin}
	}
	\subfloat[Poisson]{%
			\includegraphics[width=0.23\columnwidth]{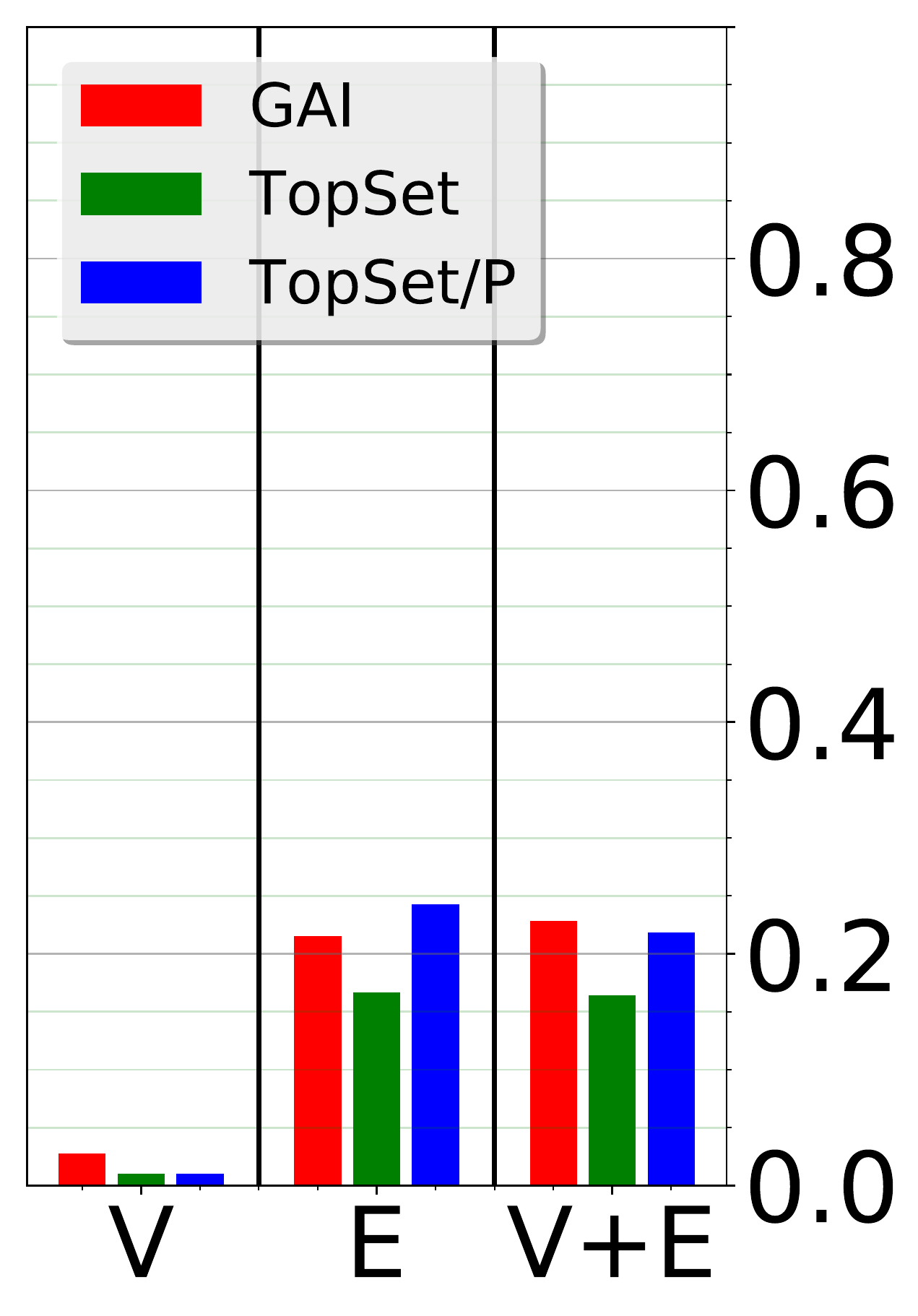}%
			\label{fig:workload4:stab_Violin}
		}
\vspace{-0.05in}
	\caption{\emph{Average of Total Stabilization time} $\overline{\psi_{t+1}}$ over $100$ control intervals for different placement strategies ($\eta=1~sec$)}
	\label{fig:workload:Mean_stabilizationTime}
\vspace{-0.15in}
\end{figure}
The migrations also impact \textbf{\emph{stabilization time}}, and in particular, the peak number of migrations on a single resource that can cause queries on it to buffer longer and take longer to drain. Fig.~\ref{fig:workload:Mean_stabilizationTime} shows the average time taken for all queries to drain the buffer and reach a steady state of event stream execution for the small setup. The plot for large setup is skipped for brevity. Here, we assume that the cost of each migration (and hence buffering time) is $\eta=1~sec$, but this may vary based on the CEP engine used and its bootstrap overheads. Since migrations are less frequent, the median stabilization time tends to $0$. We normally find the stabilization time to be in the range of sub-seconds when non-zero with some peaks reaching $\approx 20~secs$. We also see that typically, the stabilization time for GAI is smaller than TopSet and TopSet/P, indicating that fewer number of peak migrations on a resource happen.
For the large setup, we report that Edge and Vertex+Edge rebalance require a larger stabilization time for all workloads, compared to Vertex rebalance. There is also a correlation between the increase in stabilization time of $\approx 10\times$, taking up to $8~secs$, with $10\times$ more number of resources relative to the small setup.




\section{Related Work}
\label{sec:related}

Edge computing is gaining increasing attention~\cite{epema:edge}, and generalizes prior work on Wireless Sensor Networks (WSN)~\cite{Srivastava05}, Cloudlets, and Mobile Cloud~\cite{satya:comm:2015}. Constrained scheduling of applications to meet QoS and dynamic resource management across edge and cloud has gained attention~\cite{Shekhar2017}. Serendipity~\cite{Shi12} opportunistically exploits mobile resources within communication range for offloading parallel tasks of a program from a device, with mobile devices joining and leaving the system. Nebula~\cite{Ryden:2014} focuses on pushing data-intensive compute to geographically distributed edge devices with localized optimisations on location-aware data and computation placement, replication,
and recover. 
Others examine a related problem on scheduling dynamic independent tasks from Cloudlets to mobile \emph{ad hoc} Clouds~\cite{Li:2015}, with heuristics that are validated through simulation. They use both user-centric and system-centric metrics like average makespan, waiting time, slowdown, etc. We focus on edge devices that are part of the infrastructure where availability is not a concern but energy and compute constraints exist. We also support dynamic arrival and exit of dataflows, which are more complex than tasks, and for a streaming scenario.

Wide area distributed query processing has been examined for WSN. There, constrained motes collocated with sensors partition a query across the edge devices for online processing~\cite{Srivastava05}. Some of these look into stream processing across nodes with varying event rates and network performance, where the schedule has to be dynamically changed~\cite{Hwang:08}. Data partitioning and selective replication are also used along with temporal information in prior workloads to predict future query patterns and reduce the query span~\cite{turk:14}. We consider dynamism in application arrival but not event rates or resource availability. While issues of energy limits and network dynamism do exist in WSN, current edge devices have superior performance, and are complemented by Cloud resources for cooperative scheduling rather than an edge-only approach. We also see a richer set of dataflow applications and fast rates with IoT.


Big data platforms like Storm~\cite{peng:middleware:2015} and Flink are designed for streaming applications deployed on Cloud resources. Resource aware scheduling of such continuous dataflows considers some dynamism in application structure~\cite{charith:escience:2013}. CEP dataflows are a specialization of such fast data applications, but with standard query models rather than arbitrary user logic. As a result, our CEP dataflow scheduler has better awareness of the resource needs, such as energy and compute per event, compared to opaque user tasks.

Scheduling of scientific workflows on Cloud resources have been well studied using elastic resource allocation strategies~\cite{6972263}. Often they require multi-objective optimisation goals to meet the requirements of the workflows~\cite{NGUYEN2017}. While event analytic dataflows have a similar DAG model as workflows, they process data in a stream continuously than a batch, and hence all their queries are active all the time. The edge devices we consider also have constraints like energy that are based on deployment parameters like battery capacity. These open up novel scheduling problems beyond workflow that we tackle.

DAG scheduling is a well-studied problem with a number of heuristics that have been proposed to solve this NP-complete problem~\cite{wu2001efficient,Kwok99,MALIK2018}. Makespan is a common optimization criteria, with critical-path based approaches often considered as a form of list scheduling. Meta-heuristics like GA and ACO are also often used~\cite{Michalewicz96}. Placement of large-scale distributed applications on multiple clouds using heuristics and meta-heuristics have been well studied~\cite{Silva:2016,Canon2008}. We leverage some of these common strategies to solve an interesting and practical IoT scheduling problem across edge and Cloud.

\section{Conclusions}
\label{sec:conclusions}
We have proposed a distinctive problem of scheduling dynamic event analytic dataflows on edge and Cloud computing resources to support the emerging needs of IoT and smart city applications. We define the optimization problem for query placement with energy and compute constraints, customized to the unique needs of IoT resource deployments. We have proposed the TopSet heuristic based on a topological set ordering, along with a variant that considers side-effects on other dataflows. We also extend two prior GA heuristics from a static to a dynamic dataflow scenario. Rebalance strategies further improve upon the initial placement decision.
Our detailed simulations using real-world traces of queries and resources show that TopSet/P with Edge and Vertex rebalance is fast (sub-second), consistently offers a valid solution, has a makespan that out-performs in most cases, and has a mean stabilization time of $<8~sec$ even with $1000$ devices. GA-based solutions fail for larger setups, and are slow as well.  
\IEEEtriggeratref{25}


\bibliographystyle{IEEEtran}
\bibliography{paper}

\begin{thebibliography}{10}
\providecommand{\url}[1]{#1}
\csname url@samestyle\endcsname
\providecommand{\newblock}{\relax}
\providecommand{\bibinfo}[2]{#2}
\providecommand{\BIBentrySTDinterwordspacing}{\spaceskip=0pt\relax}
\providecommand{\BIBentryALTinterwordstretchfactor}{4}
\providecommand{\BIBentryALTinterwordspacing}{\spaceskip=\fontdimen2\font plus
\BIBentryALTinterwordstretchfactor\fontdimen3\font minus
  \fontdimen4\font\relax}
\providecommand{\BIBforeignlanguage}[2]{{%
\expandafter\ifx\csname l@#1\endcsname\relax
\typeout{** WARNING: IEEEtran.bst: No hyphenation pattern has been}%
\typeout{** loaded for the language `#1'. Using the pattern for}%
\typeout{** the default language instead.}%
\else
\language=\csname l@#1\endcsname
\fi
#2}}
\providecommand{\BIBdecl}{\relax}
\BIBdecl

\bibitem{bloem2014fourth}
J.~Bloem and et~al., ``The fourth industrial revolution,'' \emph{Things
  Tighten}, 2014.

\bibitem{simmhan:cise:2012}
Y.~Simmhan, V.~Prasanna, S.~Aman, A.~Kumbhare, R.~Liu, S.~Stevens, and Q.~Zhao,
  ``Cloud-based software platform for big data analytics in smart grids,''
  \emph{IEEE/AIP CiSE}, 2013.

\bibitem{strohbach2015towards}
M.~Strohbach, H.~Ziekow, V.~Gazis, and N.~Akiva, ``Towards a big data analytics
  framework for iot and smart city applications,'' in \emph{Modeling and
  processing for next-generation big-data technologies}, 2015.

\bibitem{cep-survey}
G.~Cugola and A.~Margara, ``Processing flows of information: From data stream
  to complex event processing,'' \emph{ACM CSUR}, 2012.

\bibitem{edgent}
``{Apache Edgent, v1.1.0},'' \url{http://edgent.apache.org/}.

\bibitem{siddhi11}
S.~Suhothayan, K.~Gajasinghe, I.~Loku~Narangoda, S.~Chaturanga, S.~Perera, and
  V.~Nanayakkara, ``Siddhi: A second look at complex event processing
  architectures,'' in \emph{ACM Gateway Comp. Env.}, 2011.

\bibitem{debs-challenge-plug}
Z.~Jerzak and H.~Ziekow, ``The debs 2014 grand challenge,'' in \emph{ACM DEBS},
  2014.

\bibitem{varshney:icfec}
P.~Varshney and Y.~Simmhan, ``Demystifying fog computing: Characterizing
  architectures, applications and abstractions,'' in \emph{IEEE ICFEC}, 2017.

\bibitem{kumar:icml:2017}
A.~Kumar, S.~Goyal, and M.~Varma, ``Resource-efficient machine learning in 2 kb
  ram for the internet of things,'' in \emph{International Conference on
  Machine Learning}, 2017, pp. 1935--1944.

\bibitem{ghosh:tcps:2017}
\BIBentryALTinterwordspacing
R.~Ghosh and Y.~Simmhan, ``Distributed scheduling of event analytics across
  edge and cloud,'' \emph{ACM Transactions on Cyber-Physical Systems (TCPS)},
  to Appear. [Online]. Available: \url{http://arxiv.org/abs/1608.01537}
\BIBentrySTDinterwordspacing

\bibitem{satya:comm:2015}
M.~Satyanarayanan, R.~Schuster, M.~Ebling, G.~Fettweis, H.~Flinck, K.~Joshi,
  and K.~Sabnani, ``An open ecosystem for mobile-cloud convergence,''
  \emph{IEEE Comm. Magazine}, 2015.

\bibitem{epema:edge}
P.~Garcia~Lopez, A.~Montresor, D.~Epema, A.~Datta, T.~Higashino, A.~Iamnitchi,
  M.~Barcellos, P.~Felber, and E.~Riviere, ``Edge-centric computing: Vision and
  challenges,'' \emph{ACM Comp. Comm. Rev.}, 2015.

\bibitem{smartx}
\BIBentryALTinterwordspacing
Smartx, ``{IISc Smart Campus: Closing the loop from Network to Knowledge},''
  2016. [Online]. Available: \url{http://smartx.cds.iisc.ac.in}
\BIBentrySTDinterwordspacing

\bibitem{amrutur:lightpole}
{B. Amrutur, et al.}, ``{An Open Smart City IoT Test Bed: Street Light Poles as
  Smart City Spines},'' in \emph{ACM/IEEE IoTDI}, 2017.

\bibitem{yannuzzi-2017}
M.~Yannuzzi, F.~van Lingen, A.~Jain, O.~L. Parellada, M.~M. Flores, D.~Carrera,
  J.~L. Perez, D.~Montero, P.~Chacin, A.~Corsaro, and A.~Olive, ``A new era for
  cities with fog computing,'' \emph{IEEE Internet Computing}, 2017.

\bibitem{mishra:iotn:2015}
P.~Misra, Y.~Simmhan, and J.~Warrior, ``Towards a practical architecture for
  internet of things: An india-centric view,'' \emph{IEEE Internet of Things
  Newsletter}, pp. 1--2, 2015.

\bibitem{who-chlorine}
``Measuring chlorine levels in water supplies,'' World Health Organization,
  Tech. Rep., 2013.

\bibitem{Kwok99}
Y.-K. Kwok and I.~Ahmad, ``Static scheduling algorithms for allocating directed
  task graphs to multiprocessors,'' \emph{ACM CSUR}, 1999.

\bibitem{zhao2006scheduling}
H.~Zhao and R.~Sakellariou, ``Scheduling multiple dags onto heterogeneous
  systems,'' in \emph{IPDPS}, 2006.

\bibitem{Michalewicz96}
Z.~Michalewicz, \emph{Genetic Algorithms + Data Structures = Evolution
  Programs. London, UK: Springer-Verlag}, 1996.

\bibitem{wu2001efficient}
M.-Y. Wu, W.~Shu, and J.~Gu, ``Efficient local search far dag scheduling,''
  \emph{IEEE Transactions on parallel and distributed systems}, 2001.

\bibitem{RTRG}
R.~A. Shafik, B.~M. Al-Hashimi, and K.~Chakrabarty, ``Soft error-aware design
  optimization of low power and time-constrained embedded systems,'' in
  \emph{DATE}, 2010.

\bibitem{Arampatzis:2008}
A.~Arampatzis and J.~Kamps, ``A study of query length,'' in \emph{ACM SIGIR},
  2008.

\bibitem{Silva:2016}
P.~Silva, C.~Perez, and F.~Desprez, ``Efficient heuristics for placing
  large-scale distributed applications on multiple clouds,'' in \emph{2016 16th
  IEEE/ACM International Symposium on Cluster, Cloud and Grid Computing
  (CCGrid)}, May 2016, pp. 483--492.

\bibitem{Srivastava05}
U.~Srivastava, K.~Munagala, and J.~Widom, ``Operator placement for in-network
  stream query processing,'' in \emph{ACM PODS}, 2005.

\bibitem{Shekhar2017}
S.~Shekhar and A.~Gokhale, ``Dynamic resource management across cloud-edge
  resources for performance-sensitive applications,'' in \emph{2017 17th
  IEEE/ACM International Symposium on Cluster, Cloud and Grid Computing
  (CCGRID)}, May 2017, pp. 707--710.

\bibitem{Shi12}
C.~Shi, V.~Lakafosis, M.~H. Ammar, and E.~W. Zegura, ``Serendipity: Enabling
  remote computing among intermittently connected mobile devices,'' in
  \emph{ACM MobiHoc}, 2012.

\bibitem{Ryden:2014}
M.~Ryden, K.~Oh, A.~Chandra, and J.~Weissman, ``Nebula: Distributed edge cloud
  for data intensive computing,'' in \emph{2014 IEEE International Conference
  on Cloud Engineering}, March 2014, pp. 57--66.

\bibitem{Li:2015}
B.~Li, Y.~Pei, H.~Wu, and B.~Shen, ``Heuristics to allocate high-performance
  cloudlets for computation offloading in mobile ad hoc clouds,'' \emph{J.
  Supercomput.}, 2015.

\bibitem{Hwang:08}
J.~H. Hwang, U.~Cetintemel, and S.~Zdonik, ``Fast and highly-available stream
  processing over wide area networks,'' in \emph{IEEE ICDE}, 2008.

\bibitem{turk:14}
A.~Turk, R.~O. Selvitopi, H.~Ferhatosmanoglu, and C.~Aykanat, ``Temporal
  workload-aware replicated partitioning for social networks,'' \emph{IEEE
  Transactions on Knowledge and Data Engineering}, 2014.

\bibitem{peng:middleware:2015}
B.~Peng, M.~Hosseini, Z.~Hong, R.~Farivar, and R.~Campbell, ``R-storm:
  Resource-aware scheduling in storm,'' in \emph{Middleware}, 2015.

\bibitem{charith:escience:2013}
C.~Wickramaarachchi and Y.~Simmhan, ``Continuous dataflow update strategies for
  mission-critical applications,'' in \emph{IEEE eScience}, 2013.

\bibitem{6972263}
R.~F. d.~Silva, W.~Chen, G.~Juve, K.~Vahi, and E.~Deelman, ``Community
  resources for enabling research in distributed scientific workflows,'' in
  \emph{IEEE e-Science}, 2014.

\bibitem{NGUYEN2017}
H.~A. Nguyen, Z.~van Iperen, S.~Raghunath, D.~Abramson, T.~Kipouros, and
  S.~Somasekharan, ``Multi-objective optimisation in scientific workflow,''
  \emph{Procedia Computer Science}, vol. 108, no. Supplement C, pp. 1443 --
  1452, 2017, international Conference on Computational Science, ICCS 2017,
  12-14 June 2017, Zurich, Switzerland.

\bibitem{MALIK2018}
A.~Malik, C.~Walker, M.~O’Sullivan, and O.~Sinnen, ``Satisfiability modulo
  theory (smt) formulation for optimal scheduling of task graphs with
  communication delay,'' \emph{Computers \& Operations Research}, vol.~89, no.
  Supplement C, pp. 113 -- 126, 2018.

\bibitem{Canon2008}
L.-C. Canon, E.~Jeannot, R.~Sakellariou, and W.~Zheng, \emph{Comparative
  Evaluation Of The Robustness Of DAG Scheduling Heuristics}.\hskip 1em plus
  0.5em minus 0.4em\relax Boston, MA: Springer US, 2008, pp. 73--84.

\end{thebibliography}

\newpage

\end{document}